%% file: main.tex
\def\buildpreprint{}
\documentclass[11pt,letterpaper]{article}
\input{preamble}

\title{Diagnosing source-nested label confounding in pooled biological data}

\author{%
  Robert Schuler\textsuperscript{1,*}\,\orcidlink{0000-0002-8956-5707} and
  Carol Schuler\textsuperscript{2}\,\orcidlink{0009-0002-5343-434X}\\[0.6em]
  {\small \textsuperscript{1}\emph{Information Sciences Institute, University of Southern California, Marina del Rey, California, USA}}\\
  {\small \textsuperscript{2}\emph{Creatuor.com, Redondo Beach, California, USA}}
}
\date{}

\begin{document}
\maketitle
\begingroup
\renewcommand{\thefootnote}{*}%
\footnotetext{Correspondence: \href{mailto:schuler@isi.edu}{schuler@isi.edu}}%
\ifpreprint
\renewcommand{\thefootnote}{}%
\footnotetext{Preprint. This work has not been peer reviewed.}%
\fi
\endgroup

\input{sections/00_summary}
\input{sections/01_bigger_picture}
\input{sections/00b_highlights}

\input{sections/02_introduction}
\input{sections/03_results}
\input{sections/04_discussion}
\input{sections/05_methods}
\input{sections/06_backmatter}

\bibliographystyle{unsrtnat}
\bibliography{references}

\input{sections/08_document_s1}

\end{document}

%% file: preamble.tex
\newif\ifpreprint
\ifdefined\buildpreprint \preprinttrue \fi
\newif\ifpreprintpubliccode
\newcommand{\preprinttag}{v1.0-preprint}

\usepackage[utf8]{inputenc}
\usepackage[T1]{fontenc}
\usepackage[expansion=false]{microtype}
\usepackage[letterpaper,margin=1in]{geometry}
\usepackage{setspace}
\usepackage{amsmath,amssymb}
\usepackage{textcomp}

\usepackage{graphicx}
\graphicspath{{figures/}}
\usepackage{booktabs}
\usepackage{array}
\usepackage{longtable}
\usepackage[font=small,labelfont=bf,labelsep=period]{caption}
\usepackage{float}

\usepackage{enumitem}
\setlist[itemize]{leftmargin=*}

\usepackage[super,sort&compress]{natbib}
\usepackage{xcolor}
\definecolor{linkblue}{RGB}{0,51,153}
\usepackage[colorlinks=true,linkcolor=linkblue,citecolor=linkblue,urlcolor=linkblue]{hyperref}
\usepackage{orcidlink} 

\usepackage{titlesec}
\titleformat{\section}{\large\bfseries}{}{0em}{}
\titleformat{\subsection}{\normalsize\bfseries}{}{0em}{}
\titleformat{\subsubsection}{\normalsize\itshape\bfseries}{}{0em}{}
\titlespacing*{\section}{0pt}{1.4em}{0.6em}
\titlespacing*{\subsection}{0pt}{1.0em}{0.4em}
\titlespacing*{\subsubsection}{0pt}{0.8em}{0.3em}

\usepackage[most]{tcolorbox}
\newtcolorbox{pbox}[1]{colback=gray!6,colframe=gray!60,fonttitle=\bfseries,title=#1,boxrule=0.5pt,arc=1pt,left=6pt,right=6pt,top=4pt,bottom=4pt}

\usepackage[nameinlink,noabbrev]{cleveref}
\crefname{figure}{Figure}{Figures}
\crefname{table}{Table}{Tables}

\newcommand{\crefpanels}[3]{\namecrefs{#1}~\labelcref{#1}#2 and~\labelcref{#1}#3}
\newcommand{\secref}[1]{\nameref{#1}}
\newcommand{\methodsref}[1]{\nameref{sec:methods}, \emph{\nameref{#1}}}

\newcommand{\sidef}[6]{%
  \expandafter\def\csname si@kind@#1\endcsname{#2}%
  \expandafter\def\csname si@kinds@#1\endcsname{#3}%
  \expandafter\def\csname si@num@#1\endcsname{#4}%
  \expandafter\def\csname si@sheet@#1\endcsname{#5}%
  \expandafter\def\csname si@tgt@#1\endcsname{#6}}
\newcommand{\sicheck}[1]{\ifcsname si@num@#1\endcsname\else
  \PackageError{si-refs}{Unknown supplemental ID #1 (not in SI/SI_manifest.csv)}{}\fi}
\newcommand{\sinum}[1]{\sicheck{#1}\csname si@num@#1\endcsname\csname si@sheet@#1\endcsname}
\newcommand{\silink}[2]{\sicheck{#1}\edef\sitarget{\csname si@tgt@#1\endcsname}\hyperref[\sitarget]{#2}}
\newcommand{\siname}[1]{\sicheck{#1}\csname si@kind@#1\endcsname~\sinum{#1}}
\newcommand{\siref}[1]{\silink{#1}{\siname{#1}}}
\newcommand{\sirefs}[2]{\sicheck{#1}\csname si@kinds@#1\endcsname~\silink{#1}{\sinum{#1}} and~\silink{#2}{\sinum{#2}}}
\newcommand{\sirange}[2]{\sicheck{#1}\csname si@kinds@#1\endcsname~\silink{#1}{\sinum{#1}}--\silink{#2}{\sinum{#2}}}
\input{si_refs}

%% file: si_refs.tex
\sidef{MS-SI-DOCUMENT}{Document}{Documents}{S1}{}{text:MS-SI-DOCUMENT}
\sidef{MS-SI-WB-STAGE2-PREDICTION-PROVENANCE}{Table}{Tables}{S1}{}{tab:MS-SI-WB-STAGE2-PREDICTION-PROVENANCE}
\sidef{MS-SI-TAB-STAGE2-ERRORS-COVARIATES}{Table}{Tables}{S1}{, sheet~A}{tab:MS-SI-WB-STAGE2-PREDICTION-PROVENANCE}
\sidef{MS-SI-TAB-STAGE2-STUDY-DECOMP}{Table}{Tables}{S1}{, sheet~B}{tab:MS-SI-WB-STAGE2-PREDICTION-PROVENANCE}
\sidef{MS-SI-TAB-WITHIN-STUDY-CORRECTNESS}{Table}{Tables}{S1}{, sheet~C}{tab:MS-SI-WB-STAGE2-PREDICTION-PROVENANCE}
\sidef{MS-SI-TAB-ENTROPY-BY-OUTCOME}{Table}{Tables}{S1}{, sheet~D}{tab:MS-SI-WB-STAGE2-PREDICTION-PROVENANCE}
\sidef{MS-SI-TAB-CONDITIONAL-ATTRIBUTION}{Table}{Tables}{S1}{, sheet~E}{tab:MS-SI-WB-STAGE2-PREDICTION-PROVENANCE}
\sidef{MS-TAB-STAGE2-PERCLASS}{Table}{Tables}{S1}{, sheet~F}{tab:MS-SI-WB-STAGE2-PREDICTION-PROVENANCE}
\sidef{MS-SI-TAB-ERROR-PROVENANCE}{Table}{Tables}{S1}{, sheet~G}{tab:MS-SI-WB-STAGE2-PREDICTION-PROVENANCE}
\sidef{MS-SI-TAB-LABEL-PAIRS}{Table}{Tables}{S1}{, sheet~H}{tab:MS-SI-WB-STAGE2-PREDICTION-PROVENANCE}
\sidef{MS-TAB-STAGE2-DECOMP}{Table}{Tables}{S1}{, sheet~I}{tab:MS-SI-WB-STAGE2-PREDICTION-PROVENANCE}
\sidef{MS-SI-WB-INTERVALS-CALIBRATION}{Table}{Tables}{S2}{}{tab:MS-SI-WB-INTERVALS-CALIBRATION}
\sidef{MS-SI-TAB-HEADLINE-CIS}{Table}{Tables}{S2}{, sheet~A}{tab:MS-SI-WB-INTERVALS-CALIBRATION}
\sidef{MS-TAB-CALIBRATION}{Table}{Tables}{S2}{, sheet~B}{tab:MS-SI-WB-INTERVALS-CALIBRATION}
\sidef{MS-TAB-HEADLINE}{Table}{Tables}{S2}{, sheet~C}{tab:MS-SI-WB-INTERVALS-CALIBRATION}
\sidef{MS-TAB-PROVENANCE-INFO}{Table}{Tables}{S2}{, sheet~D}{tab:MS-SI-WB-INTERVALS-CALIBRATION}
\sidef{MS-SI-WB-STAGE2-SPLIT-PROVENANCE}{Table}{Tables}{S3}{}{tab:MS-SI-WB-STAGE2-SPLIT-PROVENANCE}
\sidef{MS-SI-TAB-LABEL-BY-STUDY}{Table}{Tables}{S3}{, sheet~A}{tab:MS-SI-WB-STAGE2-SPLIT-PROVENANCE}
\sidef{MS-SI-TAB-PER-LABEL-PROVENANCE}{Table}{Tables}{S3}{, sheet~B}{tab:MS-SI-WB-STAGE2-SPLIT-PROVENANCE}
\sidef{MS-SI-TAB-LABEL-BY-COVARIATE}{Table}{Tables}{S3}{, sheet~C}{tab:MS-SI-WB-STAGE2-SPLIT-PROVENANCE}
\sidef{MS-SI-TAB-STAGE2-INFORMATION}{Table}{Tables}{S3}{, sheet~D}{tab:MS-SI-WB-STAGE2-SPLIT-PROVENANCE}
\sidef{MS-SI-TAB-ENHANCER-STRAIN-GROUPS}{Table}{Tables}{S3}{, sheet~E}{tab:MS-SI-WB-STAGE2-SPLIT-PROVENANCE}
\sidef{MS-SI-WB-STAGE1-SPLIT-PROVENANCE}{Table}{Tables}{S4}{}{tab:MS-SI-WB-STAGE1-SPLIT-PROVENANCE}
\sidef{MS-SI-TAB-STAGE1-STUDY-BY-SPLIT}{Table}{Tables}{S4}{, sheet~A}{tab:MS-SI-WB-STAGE1-SPLIT-PROVENANCE}
\sidef{MS-SI-TAB-ADULT-POOL-BY-STUDY}{Table}{Tables}{S4}{, sheet~B}{tab:MS-SI-WB-STAGE1-SPLIT-PROVENANCE}
\sidef{MS-SI-TAB-STAGE1-STRAIN-ZYGOSITY-INFO}{Table}{Tables}{S4}{, sheet~C}{tab:MS-SI-WB-STAGE1-SPLIT-PROVENANCE}
\sidef{MS-SI-WB-STAGE1-PREDICTION-PROVENANCE}{Table}{Tables}{S5}{}{tab:MS-SI-WB-STAGE1-PREDICTION-PROVENANCE}
\sidef{MS-SI-TAB-STAGE1-FALSE-POSITIVES}{Table}{Tables}{S5}{, sheet~A}{tab:MS-SI-WB-STAGE1-PREDICTION-PROVENANCE}
\sidef{MS-SI-TAB-STAGE1-FALSE-NEGATIVES}{Table}{Tables}{S5}{, sheet~B}{tab:MS-SI-WB-STAGE1-PREDICTION-PROVENANCE}
\sidef{MS-SI-TAB-STAGE1-PROVENANCE-SUBSETS}{Table}{Tables}{S5}{, sheet~C}{tab:MS-SI-WB-STAGE1-PREDICTION-PROVENANCE}
\sidef{MS-TAB-STAGE1-SUBSETS}{Table}{Tables}{S5}{, sheet~D}{tab:MS-SI-WB-STAGE1-PREDICTION-PROVENANCE}
\sidef{MS-SI-TAB-THRESHOLD-SWEEP}{Table}{Tables}{S5}{, sheet~E}{tab:MS-SI-WB-STAGE1-PREDICTION-PROVENANCE}
\sidef{MS-SI-WB-BASELINES}{Table}{Tables}{S6}{}{tab:MS-SI-WB-BASELINES}
\sidef{MS-SI-TAB-BASELINES}{Table}{Tables}{S6}{, sheet~A}{tab:MS-SI-WB-BASELINES}
\sidef{MS-SI-TAB-MULTISTRUCTURE-BASELINES}{Table}{Tables}{S6}{, sheet~B}{tab:MS-SI-WB-BASELINES}
\sidef{MS-TAB-BASELINES}{Table}{Tables}{S6}{, sheet~C}{tab:MS-SI-WB-BASELINES}
\sidef{MS-SI-WB-MATCHED-COHORT}{Table}{Tables}{S7}{}{tab:MS-SI-WB-MATCHED-COHORT}
\sidef{MS-SI-TAB-MATCHED-COHORT}{Table}{Tables}{S7}{, sheet~A}{tab:MS-SI-WB-MATCHED-COHORT}
\sidef{MS-SI-TAB-MATCHED-CNN}{Table}{Tables}{S7}{, sheet~B}{tab:MS-SI-WB-MATCHED-COHORT}
\sidef{MS-TAB-MATCHED}{Table}{Tables}{S7}{, sheet~C}{tab:MS-SI-WB-MATCHED-COHORT}
\sidef{MS-TAB-MATCHED-PAIRED}{Table}{Tables}{S7}{, sheet~D}{tab:MS-SI-WB-MATCHED-COHORT}

%% file: sections/00_summary.tex
\ifpreprint\else
\section*{Summary}
Pooling biological data from independent studies gives machine learning larger, more diverse datasets. Each study, however, examines only a few conditions under its own protocols, so a sample's source largely determines its label and a model can predict labels without learning biology. We investigate this problem in pooled micro-CT scans of mouse skulls, training deep-learning classifiers to distinguish mutants from controls and to identify specific mutations. Both reached about 85\% accuracy, and many errors confused mutations that contributing studies had examined together, suggesting the classifiers had learned meaningful morphology. Checking predictions against each specimen's study and genetic background showed that provenance alone can reproduce most of this accuracy and that nearly all errors stay within them, leaving modest and uncertain evidence of mutant-specific morphology. We call this hazard source-nested label confounding and provide six checks that expose it from provenance metadata and model predictions before biological interpretation.

\medskip
\noindent\textbf{Keywords:} shortcut learning, confounding, provenance, batch effects, pooled data, micro-CT, craniofacial morphology, 3D convolutional neural network, geometric morphometrics, model evaluation
\fi

%% file: sections/01_bigger_picture.tex
\newcommand{\bigpicturetext}{%
Scientific collections increasingly pool data from independent studies, giving machine learning far more samples than any single study provides. Pooling, however, introduces a hazard. Contributing studies differ in many experimental or observational factors, including genetic background and acquisition protocol, and each typically examines only one or a few conditions, such as a gene mutation, a disease, or a treatment. A model that learns to recognize \emph{where} a sample came from can then predict \emph{what} it is without learning the biology. Accuracy and other metrics may suggest strong performance, and the model's errors can masquerade as biological insight.

We demonstrate this with deep-learning classifiers trained on thousands of pooled micro-CT scans of mouse skulls. Landmark-based shape analysis shows the same pattern, indicating that the problem is inherent to how the data were collected and combined rather than to any particular method. We call this hazard \emph{source-nested label confounding}. Changing the model alone cannot resolve it, because where labels are nested within sources, the data cannot show whether a prediction reflects the biology or the source. Instead, we recommend six diagnostic checks as an audit that requires only provenance metadata and model predictions. Together, these checks show when apparent performance reflects a sample's source and how much remains once the source is held fixed. The same nesting of labels within sites, cohorts, or batches can arise in other pooled data, such as radiographs, multi-site neuroimaging, and genomics. Beyond diagnosis, we encourage curators of pooled collections to publish provenance summary tables alongside their data, making source--label overlap visible before further analysis.%
}
\ifpreprint
\section*{Abstract}
\bigpicturetext
\else
\begin{pbox}{The bigger picture}
\bigpicturetext
\end{pbox}
\fi

%% file: sections/00b_highlights.tex
\ifpreprint\else
\begin{pbox}{Highlights and eTOC blurb (uploaded as a separate document at submission)}
\textbf{Highlights}
\begin{itemize}[nosep]
\item In pooled biological data, a sample's source can largely determine its label
\item Study and strain alone reach most of the classifiers' accuracy on pooled mouse skulls
\item Deep learning and landmark shape analysis are misled alike by the pooled design
\item Six checks using provenance metadata expose source-nested label confounding
\end{itemize}

\textbf{eTOC blurb.} Pooling biological data can nest labels within sources, so classifiers can recognize a sample's source instead of its biology. Schuler and Schuler show this source-nested label confounding in mouse skull micro-CT data with deep-learning and landmark classifiers and present six diagnostic checks that detect it from provenance metadata.
\end{pbox}
\fi

%% file: sections/02_introduction.tex
\section{Introduction}\label{sec:introduction}

Machine learning in biomedical research increasingly relies on pooled multi-study collections to assemble larger and more diverse samples than any single study provides. Craniofacial morphology illustrates both the appeal and the risk of this approach. Skull shape is highly sensitive to genetic perturbation,\cite{wilkie2001} and mutations affecting craniofacial development may produce subtle, localized effects that sparse sets of hand-placed landmarks can miss.\cite{maga2017,toussaint2021} Large-scale mouse phenotyping consortia now characterize thousands of knockout lines,\cite{dickinson2016} and open repositories such as FaceBase\cite{schuler2022facebase,samuels2020facebase} distribute pooled collections of micro-CT specimens contributed by many independent studies.\cite{devine2022} Landmark-free volumetric pipelines\cite{toussaint2021} and deep-learning classifiers capture shape directly from volumes, yet pooling independent studies to train them introduces a hazard that conventional evaluation can miss.

Any classifier may exploit features that correlate with the label, and one that learns directly from images has the greatest latitude to do so. Shortcut learning is well documented in natural images\cite{geirhos2020} and medical imaging, where models have inferred hospital of origin from radiographs,\cite{zech2018} exploited acquisition artifacts rather than pathology,\cite{degrave2021} and relied on hospital-process variables rather than diagnostic features.\cite{badgeley2019} Multi-site neuroimaging similarly contains enough site-specific structure for scans to be assigned to their source study well above chance,\cite{wachinger2021} while batch effects remain a long-recognized hazard across high-throughput biology.\cite{leek2010,johnson2007} Standard evaluation tools address many of these risks through grouped or blocked cross-validation,\cite{roberts2017cv,saeb2017} explicit checks for data leakage,\cite{kaufman2012,kapoor2023} and confounder-aware training and evaluation.\cite{zhao2020,badgeley2019} Confounding of the label with batch or source is itself well documented. It biases cross-validated performance estimates even after batch correction,\cite{soneson2014} has been described as confounding by provenance in multi-institution clinical text classification,\cite{ding2023} and arises among nested experimental factors in cell imaging.\cite{rogozhnikov2022} We address the configuration in which this confounding is built into the pooled design itself: most labels come from only one or a few contributing sources (studies, sites, or laboratories), and within a source, labels may be further separated by biological background such as strain. Holding out a source then removes some labels entirely, and because sources often study closely linked conditions, shortcut errors resemble biological relationships. We call this configuration \emph{source-nested label confounding}.

A model trained on a pooled resource may exploit four kinds of signal, which we distinguish throughout. Three arise from provenance and typically differ between sources: (i) \emph{technical acquisition}---scanner or platform, protocol, reconstruction, fixation, batch, and operator; (ii) \emph{study design}---how a source selected or generated its samples, such as colony, allele construction, and treatment groups in an experiment, or recruitment site and ascertainment of affected and unaffected individuals in a clinical cohort; and (iii) \emph{biological background}---genetic background (strain in mouse, ancestry in humans), sex, age, and zygosity, which are biological variables but not the ones the label denotes. A sample's source therefore stands in for all three, including any that are not recorded. The fourth, (iv) the \emph{target phenotype}, is the label-associated phenotypic difference the model is intended to learn, here mutation- or line-associated morphology. Only the target phenotype is the intended signal, and only technical acquisition is a pure artifact in the traditional sense.

Differences in study design and biological background are not artifacts, but pooling makes them predict the label. A model that relies on them makes errors among the related conditions each study examined, which mimic biological similarity. Because most labels are nested within sources, standard accuracy metrics cannot separate the target phenotype from provenance. A valid evaluation must therefore measure performance among samples from the same source and background.

Here we show that conventional evaluation overlooks this confounding and present a diagnostic framework to expose it. We trained two deep-learning classifiers on pooled micro-CT volumes: one distinguishing mutants from controls and another identifying specific mutations. Both achieved roughly 85\% test accuracy, and about half of the mutation classifier's errors confused related mutations that a contributing study had examined together,\cite{percival2017,hirsch2018,hirsch2022} which could be interpreted as evidence that the classifiers had learned meaningful morphological features. However, auditing predictions against provenance metadata revealed that a specimen's study and genetic background alone can reproduce most of this accuracy, and that nearly all errors confused labels that share both. A study often examines related genes together, so the biological plausibility of the errors follows from provenance. A landmark-based classifier showed a similar pattern, indicating that the issue stems from the pooled design rather than the classification method. Comparing samples only where both are held fixed left modest, uncertain evidence of mutant-specific morphology. To help researchers interpret classifiers trained on pooled resources with confidence, we provide six reusable diagnostic checks that use minimal provenance metadata and model predictions to expose source-nested label confounding before biological interpretation.

%% file: sections/03_results.tex
\section{Results}\label{sec:results}

\subsection{Pooled data and classification tasks}\label{sec:data-and-tasks}

Our data come from 5,814 adult mouse skull micro-CT specimens pooled from 23 contributing studies in MusMorph\cite{devine2022,musmorph_project} (\methodsref{sec:data-source-and-specimen-selection}). Some studies contribute both mutants and their own controls, others contribute only mutants, and four strain panels surveying natural genetic variation contribute only non-mutant mice, labeled as controls. MusMorph's adult category begins at postnatal day 14 and so includes juveniles, and its mutant group includes the Longshanks selected lines, which have no engineered mutation. MusMorph records each specimen's contributing study, background strain, genotype, zygosity, treatment, sex, and life phase, but not litter, scanner, acquisition session, or batch.

We framed the analysis as a hypothetical two-stage pipeline: the Stage 1 \emph{control-versus-mutant classifier} separates controls from mutants, and the Stage 2 \emph{gene-label classifier} assigns each mutant to one of 16 gene labels---the gene or genetic element MusMorph records for that mutant (\methodsref{sec:classification-targets-and-provenance-fields}). For each stage, we trained a three-dimensional convolutional neural network (3D CNN) on image volumes alone, without metadata inputs (\cref{fig:MS-FIG-ARCHITECTURE}; \methodsref{sec:network-training}). The network is a standard 3D ResNet-18 applied to downsampled volumes and represents a conventional workflow to be audited. Specimens were assigned to training, validation, and test partitions at random, stratified by sex, life phase and, for mutants, gene label (\methodsref{sec:splits-and-dependence-control}). Stage 1 was evaluated on a balanced test set of 422 specimens and Stage 2 on 132 test mutants, with each model trained and evaluated independently (\siref{MS-SI-TAB-CONDITIONAL-ATTRIBUTION} gives their joint performance on the shared test specimens).

\subsection{Conventional evaluation implies strong classification performance}\label{sec:conventional-evaluation}

\begin{figure}[htbp]
\centering
\includegraphics[width=\textwidth]{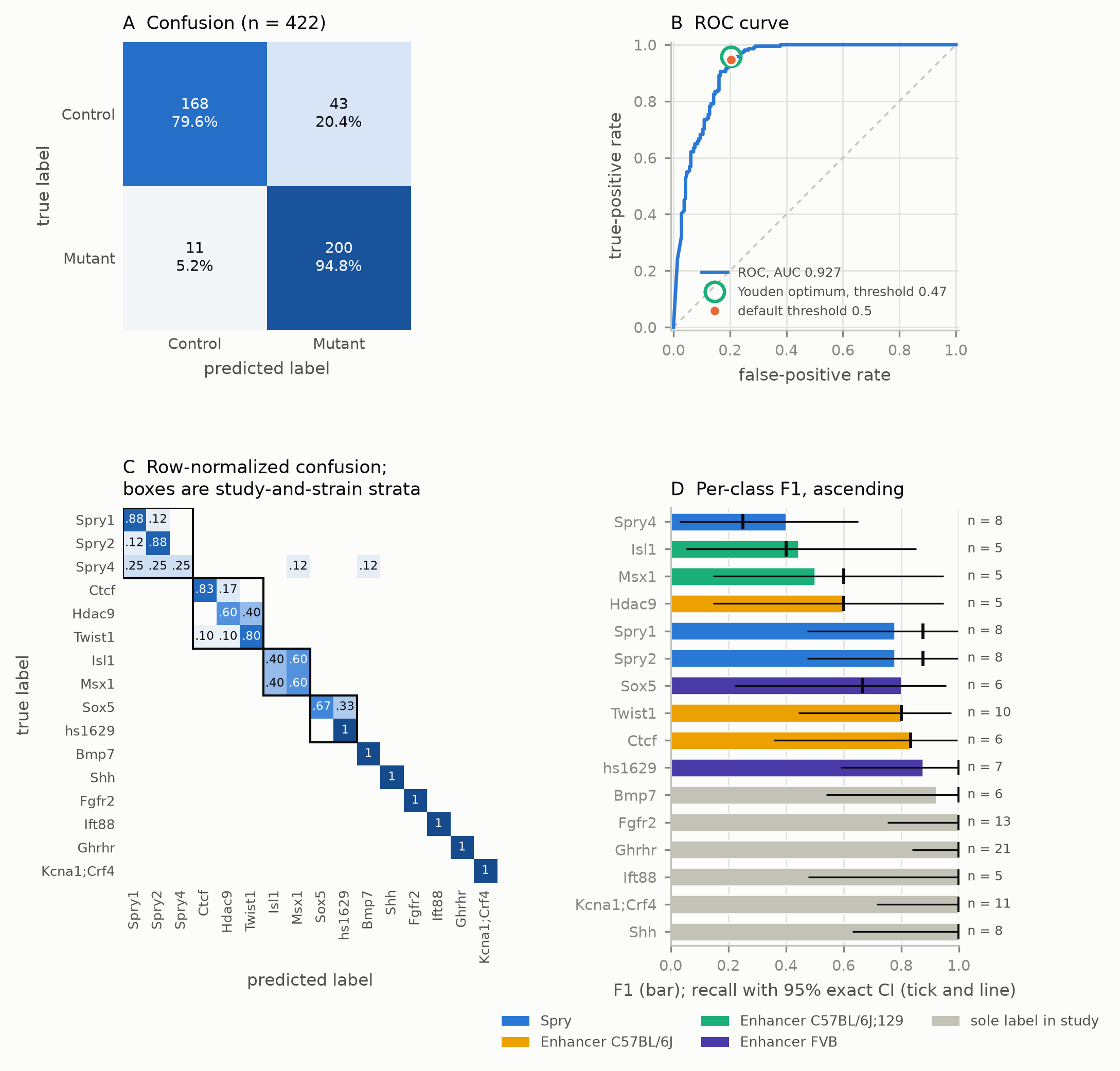}
\caption{Conventional evaluation of both stages on held-out test sets. (A) Stage 1 (control versus mutant, $n = 422$) confusion matrix (raw counts and row-normalized percentages), highlighting 43 controls misclassified as mutant versus 11 mutants misclassified as control. (B) Stage 1 ROC curve (AUC = 0.927, 95\% bootstrap CI [0.900, 0.951]); Youden-optimal ($t = 0.47$) and default 0.50 decision thresholds nearly coincide. (C) Stage 2 (16-class gene-label classification, $n = 132$) row-normalized confusion matrix, with labels ordered by contributing study and background strain. Boxes enclose the four study-and-strain strata that hold more than one label. (D) Per-class F1 scores sorted in ascending order, with recall and Clopper--Pearson 95\% CIs overlaid; bar colors indicate the stratum. Notably, most false positives in (A) are controls from studies with mutants (\cref{fig:MS-FIG-STAGE1-PROVENANCE}), and nearly all off-diagonal mass in (C) lies within the boxes (\cref{fig:MS-FIG-PROVENANCE-STRUCTURE}). Reliability analysis for Stage 1 is provided in \cref{fig:MS-SI-FIG-RELIABILITY}.}
\label{fig:MS-FIG-CONVENTIONAL}
\end{figure}

Evaluated conventionally, both stages achieved strong classification performance (\cref{fig:MS-FIG-CONVENTIONAL}; \siref{MS-SI-WB-INTERVALS-CALIBRATION}). Stage 1 reached an accuracy of 0.872 (95\% bootstrap confidence interval [0.841, 0.903]) and an ROC-AUC of 0.927. Stage 2 reached a top-1 accuracy of 0.848 [0.803, 0.894] and a macro one-vs-rest ROC-AUC of 0.962. However, a standard evaluation reveals two notable irregularities that foreshadow systemic provenance confounding. First, Stage 1 errors were severely asymmetric and overconfident. Nearly four times as many controls were misclassified as mutant as vice versa (43 vs.\ 11), and most misclassified controls received a mutant probability exceeding 0.99, rendering threshold adjustment ineffective (\cref{fig:MS-SI-FIG-THRESHOLD-SWEEP}). The model was also overconfident at every confidence level (\cref{fig:MS-SI-FIG-RELIABILITY}). Second, Stage 2 performance varied drastically across classes, ranging from perfect classification in five labels to F1 scores $\le 0.60$ for \emph{Spry4}, \emph{Isl1}, \emph{Msx1}, and \emph{Hdac9} (\cref{fig:MS-FIG-CONVENTIONAL}D; \siref{MS-SI-WB-STAGE2-PREDICTION-PROVENANCE}).

The gene-label classifier's errors also appeared to reflect meaningful biological structure. Six of the 20 Stage 2 errors were among the three Sprouty genes, members of one gene family,\cite{percival2017} and five were among three labels associated with a single regulatory locus.\cite{hirsch2018,hirsch2022} Errors of this kind could be interpreted as evidence that the model had learned morphological features shared by the mutants it confused. As the following sections show, however, the same pattern is expected from a classifier that recognizes only the contributing study and background strain of each specimen.

\subsection{Auditing classifier predictions against sample provenance}\label{sec:provenance-audit}

The following subsections audit the model predictions against provenance, forming the six diagnostics described in the \secref{sec:discussion} (\emph{\nameref{sec:practical-diagnostics-and-implications-for}}). Key audits were repeated on the validation set (\sirefs{MS-SI-WB-STAGE2-PREDICTION-PROVENANCE}{MS-SI-WB-INTERVALS-CALIBRATION}), which was used for model selection and comes from the same studies as the test set, so agreement shows consistency across splits rather than independent validation.

\begin{figure}[htbp]
\centering
\includegraphics[width=0.95\textwidth]{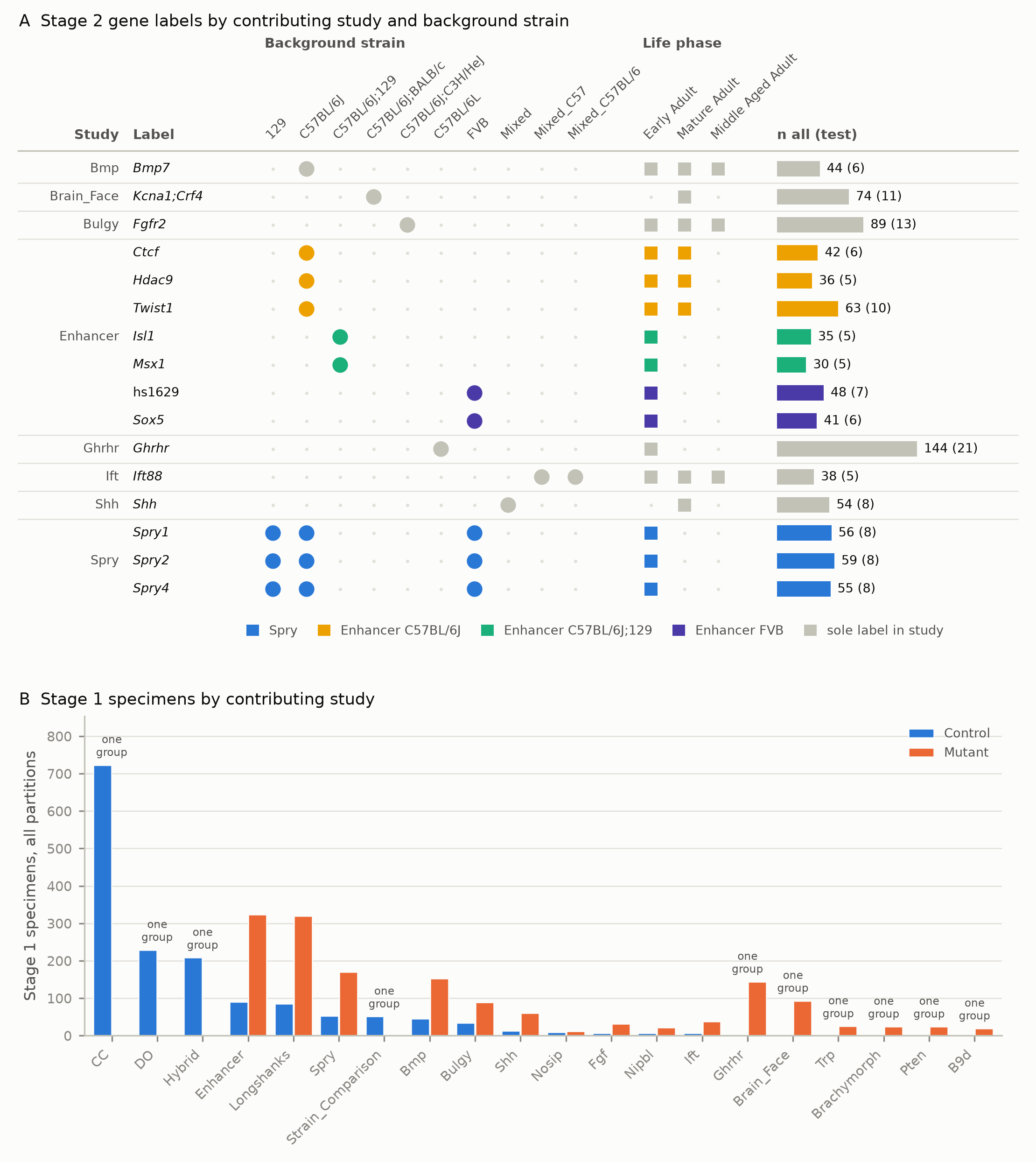}
\caption{Study and strain structure of the training data. (A) The 16 gene labels (rows), grouped by contributing MusMorph study. Circles mark the background strains and squares the life phases in which a label has specimens. Bars give specimen counts over all partitions, with test-set counts in parentheses. Each label comes from exactly one study, and six are the sole label in their study (gray). Enhancer labels are partitioned by background strain into three groups, whereas each Spry label spans the same three backgrounds. Colors mark these four study-and-strain strata as in \cref{fig:MS-FIG-CONVENTIONAL}D. Strain names are as recorded, and a semicolon denotes a single mixed background. Life phases also differ among studies but are shared by the labels within a stratum. (B) Control-versus-mutant specimen counts by contributing study across all partitions (studies with fewer than 10 specimens omitted). Four strain panels contribute only controls and mutant-only studies only mutants; studies with both groups contribute roughly one control per four mutants.}
\label{fig:MS-FIG-PROVENANCE-STRUCTURE}
\end{figure}

\subsubsection{Sample provenance alone predicts most gene labels}\label{sec:provenance-alone-predicts-most-labels}

We first examined how well provenance alone predicts gene labels. Each of the 16 gene labels originates from only one of eight contributing studies. Six are the sole label in their study, so any feature identifying the study identifies the label. The remaining ten labels derive from two multi-label studies. The Spry study contributes three Sprouty genes, each recorded on the same three background strains, while the Enhancer study contributes seven labels, each recorded on one of three background strains. Consequently, the only labels a specimen can be mistaken for without leaving its study and strain are the one or two others recorded there (\cref{fig:MS-FIG-PROVENANCE-STRUCTURE}A). Life phase, zygosity, and treatment also differ among studies, and even within fine strata, secondary covariates remain unbalanced. In the Enhancer C57BL/6J stratum, for instance, zygosity separates labels almost completely while sex distribution is skewed (\cref{fig:MS-SI-FIG-STAGE2-COMPOSITION}; \siref{MS-SI-WB-STAGE2-SPLIT-PROVENANCE}).

On the test set, knowing the study removes 69\% of label uncertainty (\methodsref{sec:study-and-strain-decomposition}). A multinomial logistic regression fit on training metadata alone (the \emph{metadata-only classifier}) achieves an accuracy of 0.720 (95\% CI [0.635, 0.794]), and predicting the majority training label for a specimen's study and strain yields similar performance (\siref{MS-SI-WB-INTERVALS-CALIBRATION}). Thus, the metadata-only classifier reaches 85\% of the CNN's overall accuracy (0.848), demonstrating that most predictive power can be attained without examining skull morphology. This baseline, rather than random chance, represents the appropriate benchmark for evaluating CNN performance. It is a lower bound on what provenance alone can predict, not a measure of how much of the CNN's own accuracy comes from provenance.

\subsubsection{Sample provenance alone matches the control-versus-mutant classifier}\label{sec:study-alone-matches-the-mutant}

We next examined how controls and mutants are distributed across studies (\cref{fig:MS-FIG-PROVENANCE-STRUCTURE}B; \siref{MS-SI-WB-STAGE1-SPLIT-PROVENANCE}). The controls come from two kinds of study. Most come from the four strain panels (the Collaborative Cross founder strains and their F1 crosses, Diversity Outbred mice, wild-derived strains and their hybrids, and a comparison of classical inbred strains). We call these \emph{strain-panel controls}. The rest were collected by the same studies that contributed mutants, as the comparison group for those mutants. We call these \emph{within-study controls}. A third kind of study contributes mutants but no controls. We call these \emph{mutant-only studies}. Across all partitions, strain-panel controls make up 78\% of controls, and only 53 of the 211 test controls are within-study controls. A \emph{study-only classifier}, which uses only each specimen's study (\methodsref{sec:baseline-classifiers}), achieves a balanced test accuracy of 0.874 (95\% CI [0.846, 0.903]), matching the 3D CNN's performance (0.872).

\subsubsection{The control-versus-mutant classifier fails on controls from studies with mutants}\label{sec:control-mutant-within-study}

\begin{figure}[htbp]
\centering
\includegraphics[width=\textwidth]{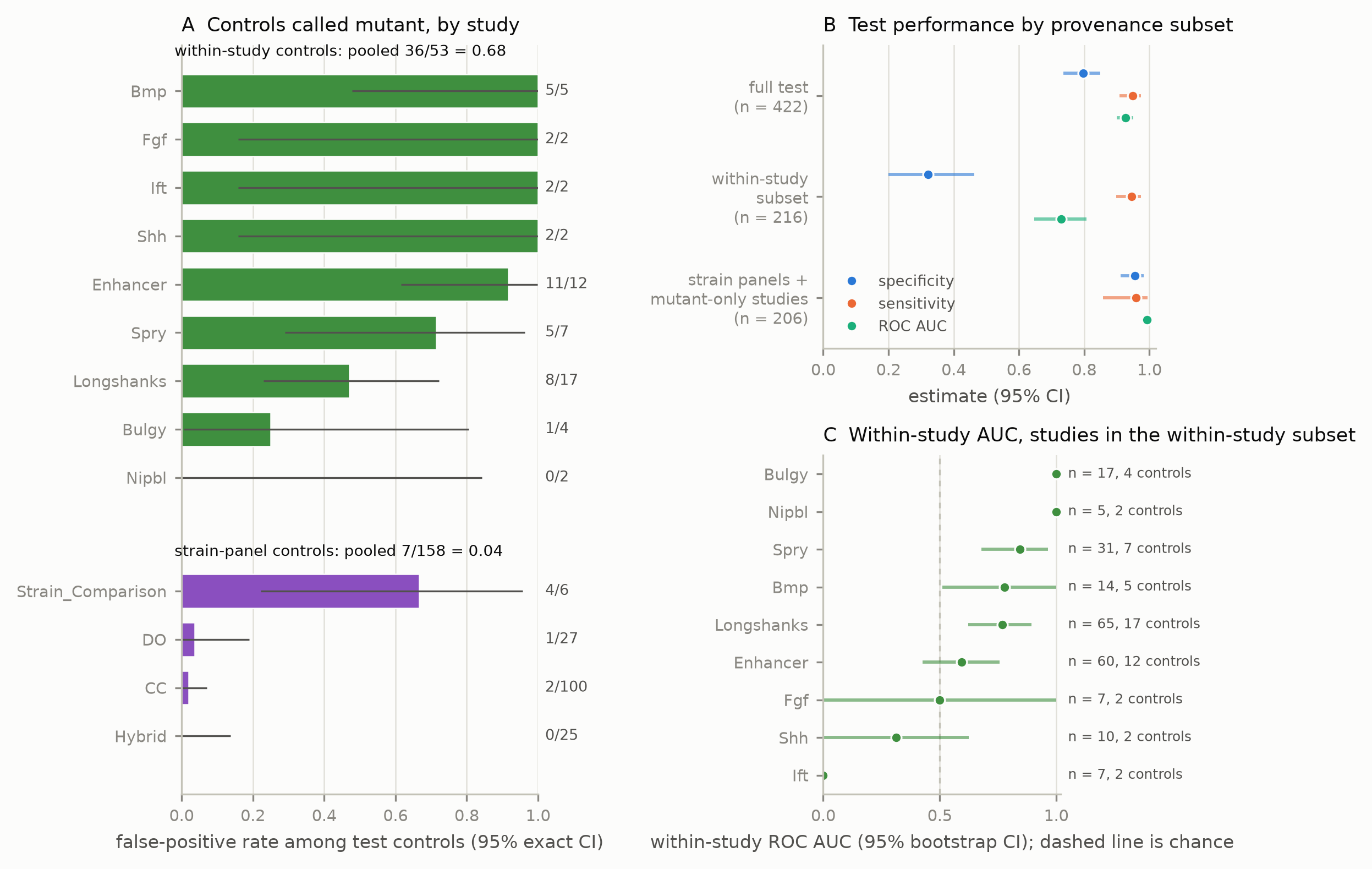}
\caption{Control-versus-mutant performance splits at the study boundary. (A) False-positive rate among test controls by contributing study, with exact 95\% intervals. The rate is 68\% (36/53) for within-study controls (upper group) and 4.4\% (7/158) for strain-panel controls (lower group). (B) Specificity, sensitivity, and ROC-AUC with 95\% intervals on the full test set, the within-study subset (nine studies pooled), and the complementary subset (strain panels plus mutant-only studies). (C) ROC-AUC within each study of the within-study subset with bootstrap intervals; most studies have two to seven test controls, so several intervals are degenerate.}
\label{fig:MS-FIG-STAGE1-PROVENANCE}
\end{figure}

We then asked whether the classifier handles controls equally well across studies, as a model capturing mutant skull morphology should. Its false positives are not distributed across controls at random (\cref{fig:MS-FIG-STAGE1-PROVENANCE}A). Only 4\% of test controls from the strain panels were misclassified as mutant, compared to 68\% of within-study controls---with most assigned a mutant probability exceeding 0.99. In fact, every within-study control on a C57BL/6J-derived background was classified as mutant; the only ones classified correctly were on CD-1 (Longshanks unselected line), FVB, and C57BL/6J;C3H/HeJ backgrounds (\siref{MS-SI-WB-STAGE1-PREDICTION-PROVENANCE}). The features driving high mutant scores are therefore shared by C57BL/6J-derived controls and mutants, which in training derive almost entirely from the same studies.

When evaluation is restricted to the 216 specimens from the nine studies contributing both test controls and test mutants (the within-study subset), specificity drops from 0.796 overall to 0.321 (95\% Clopper--Pearson CI [0.199, 0.463]) and ROC-AUC falls to 0.729 (\cref{fig:MS-FIG-STAGE1-PROVENANCE}B). Conversely, on the complementary specimens from the strain panels and mutant-only studies---where study identity alone determines the label---AUC reaches 0.992. Because this subset pools nine studies, most of its control--mutant comparisons still cross studies, so its AUC is not a within-study measure. Within individual studies, AUC varies widely and often rests on two to seven controls (\cref{fig:MS-FIG-STAGE1-PROVENANCE}C). The validation partition exhibits the same split, and adjusting the decision threshold fails to recover within-study controls because C57BL/6J controls and mutants receive the same score (\cref{fig:MS-SI-FIG-THRESHOLD-SWEEP}). Together, these findings indicate that the high overall AUC (0.927) rests mainly on comparisons across study boundaries that coincide with class labels, masking weaker, uneven discrimination within studies.

\subsubsection{Gene-label predictions and errors stay within study and strain}\label{sec:error-structure}

\begin{figure}[htbp]
\centering
\includegraphics[width=0.96\textwidth]{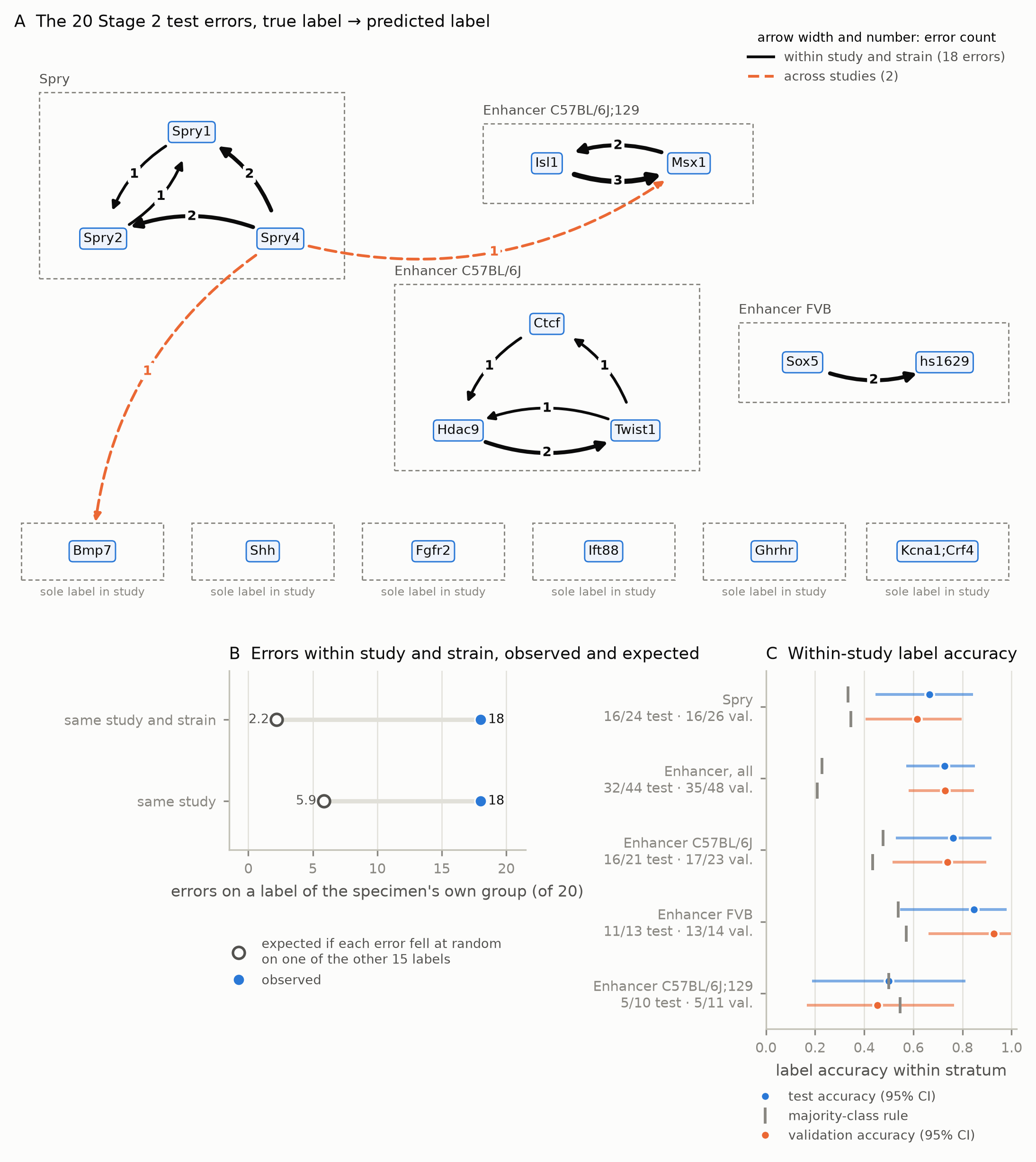}
\caption{Gene-label errors follow study and strain. (A) Directed confusion graph of the 20 test errors (true $\rightarrow$ predicted label; edge width and number give the count); boxes enclose study-and-strain strata. Eighteen errors (solid) stay within their stratum, two (dashed) cross studies. No specimen from the six sole-label studies was misclassified. (B) Number of errors whose predicted label is recorded in the specimen's own study and strain (top) or study (bottom): observed (filled) and expected if each error fell at random on one of the other 15 labels (open). (C) Label accuracy within the Spry study and within each Enhancer background strain on the test and validation sets with exact 95\% intervals, against the majority-class baseline. Accuracy exceeds the baseline everywhere except for the \emph{Isl1}/\emph{Msx1} pair.}
\label{fig:MS-FIG-STAGE2-ERRORS}
\end{figure}

To determine whether the gene-label classifier's errors reflect shared morphology among the confused labels or provenance boundaries, we first mapped each predicted label to its contributing study (\cref{fig:MS-FIG-STAGE2-ERRORS}A; \siref{MS-SI-WB-STAGE2-PREDICTION-PROVENANCE}). Across 132 test specimens, 130 predictions fell within the specimen’s true study, and on average the model placed 0.988 of its predicted probability on labels from the same study. Every specimen from the six single-label studies was classified correctly, and 18 of the 20 total errors predicted another label from the specimen's study and background strain. The other two crossed studies. Restricting predictions to a specimen's true study leaves overall accuracy unchanged. Predicting labels from the correct study does not by itself show that the network recognizes studies, because a classifier that identified mutant skull shape accurately would do the same. The network's mistakes, however, also stay within the specimen's study and background strain, indicating that they are as consistent with recognizing study and strain as with identifying mutant skull shape.

We also examined whether prediction confidence distinguishes errors within the specimen's study from the two that crossed studies. The model remained highly confident in errors that stayed within the study, so withholding the most uncertain predictions removes the two cross-study errors but keeps nearly all the others (\cref{fig:MS-FIG-STAGE2-UNCERTAINTY}). Temperature scaling fit on the validation set improved calibration at both stages without altering predictions, yet lowered the mean probability on labels from the specimen's own study only from 0.988 to 0.891 (\cref{fig:MS-SI-FIG-RELIABILITY}; \siref{MS-SI-WB-INTERVALS-CALIBRATION}). Together, these results show that calibration improves confidence estimates but leaves the classifier's predictions concentrated within the specimen's study.

\subsubsection{Per-class error patterns do not hold across partitions}\label{sec:cross-partition-patterns}

We next asked whether the direction of the errors, in which one label of a pair is mistaken for the other more often than the reverse, can be interpreted. Most gene labels have fewer than ten test specimens (12 of 16; \siref{MS-TAB-STAGE2-PERCLASS}), so such asymmetries may reflect small-sample noise rather than biological relationships. We therefore compared per-class confusions across the test and validation partitions. On the test set, \emph{Spry4} recall was substantially lower than that of \emph{Spry1} and \emph{Spry2}, superficially suggesting that it was absorbed into the other two; on the validation set, however, \emph{Spry2} was the one absorbed. Similarly, confusions between \emph{Hdac9} and \emph{Twist1}, which share a background strain, reversed direction between partitions, while \emph{Isl1} and \emph{Msx1} were distinguished no better than chance in both (\siref{MS-SI-WB-STAGE2-PREDICTION-PROVENANCE}). This instability across partitions indicates that, in classes this small, per-class error asymmetries should not be interpreted as biological similarity unless they hold in both partitions.

\subsubsection{Study design explains why the errors look biological}\label{sec:study-design-errors}

Finally, we compared where the errors fell with the labels available in each specimen's own study and strain. Of the 20 errors, 18 fell on such a label, against about 2.2 expected if each error fell at random on one of the other 15 labels (\cref{fig:MS-FIG-STAGE2-ERRORS}B; \methodsref{sec:statistical-analysis-of-error-structure}). Only eight of the 120 possible pairs of labels share a study and strain. Every one of the eight was confused at least once, whereas only two of the remaining 112 pairs were (\cref{fig:MS-FIG-STAGE2-ERRORS}A). This concentration accounts for the errors' biological plausibility, because which labels share a study is set by study design. The Spry study bred null alleles of three Sprouty genes onto the same three backgrounds to compare mutations in one pathway, and it reported grossly similar skull changes for two of them on some backgrounds.\cite{percival2017} The \emph{Twist1}, \emph{Ctcf}, and \emph{Hdac9} labels come from one group's work on a single regulatory locus.\cite{hirsch2018,hirsch2022} Given that the contributing studies reported such relationships, any confusion among these labels has a plausible biological explanation, whether or not the model uses morphology. Confusions also occurred between labels for which no comparable relationship was identified in the contributing study's documentation,\cite{musmorph_enhancer} such as \emph{Isl1} and \emph{Msx1}, which suggests that the errors follow shared provenance whether or not a relationship has been reported.

\subsection{Landmark classifiers also keep predictions within the specimen's study}\label{sec:landmark-baselines}

\begin{figure}[htbp]
\centering
\includegraphics[width=\textwidth]{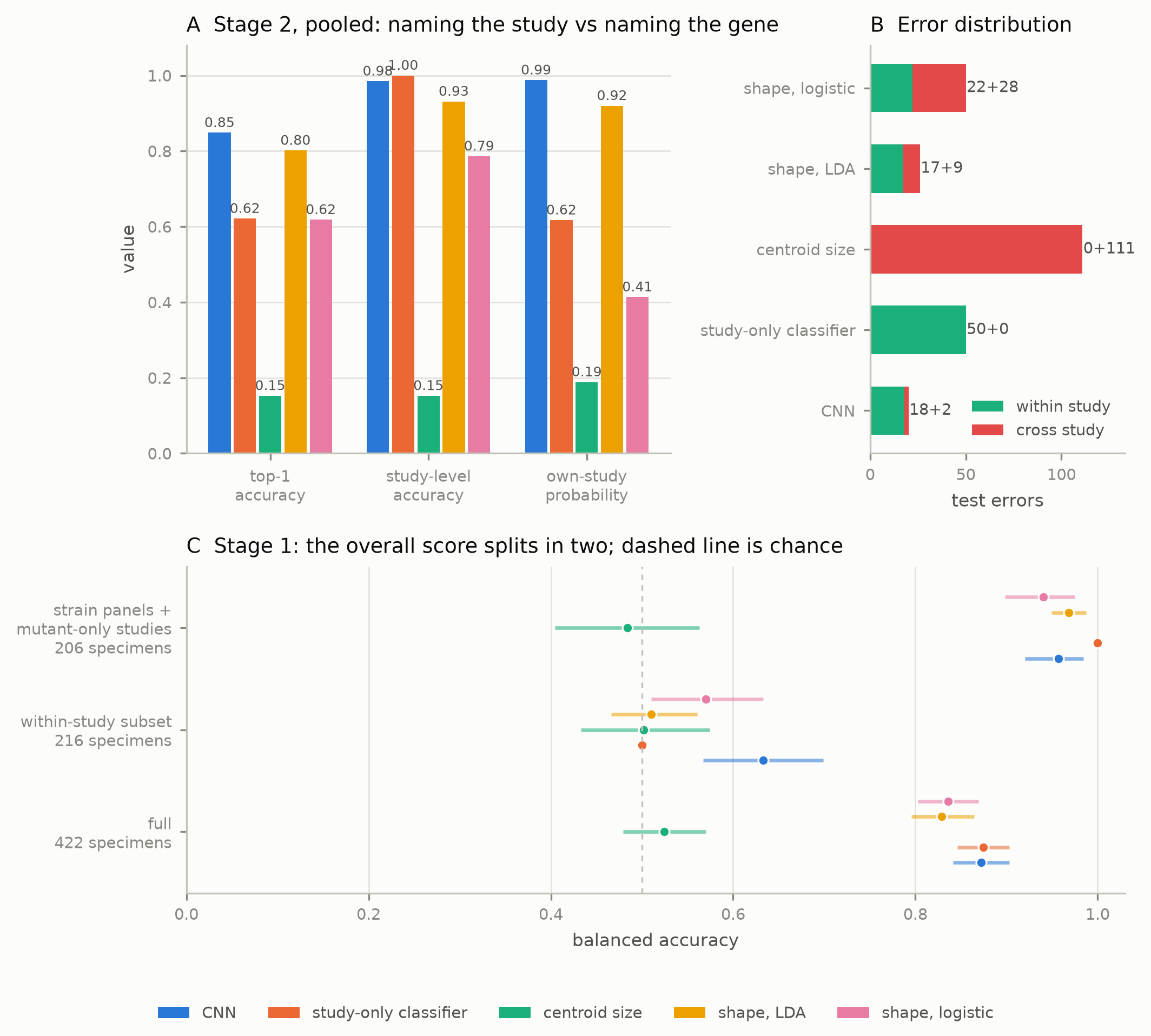}
\caption{Every classifier that can identify the study keeps its predictions within the study. (A) Stage 2 pooled top-1 accuracy, study-level accuracy (the predicted label comes from the specimen's own study), and own-study probability (the summed predicted probability on the labels of the specimen's own study, averaged over specimens) for the CNN, the study-only classifier, the centroid-size classifier (labeled ``centroid size''), and the two landmark classifiers (labeled ``shape, LDA'' and ``shape, logistic''). (B) Stage 2 test errors split by whether the predicted label lies in the specimen's true study. The landmark classifier reproduces the CNN's pattern, identifying the study far more reliably than the label, with most errors within the study. (C) Stage 1 balanced accuracy on the full test set and on the two provenance-defined halves (the within-study subset, and the strain panels plus mutant-only studies) for the same classifiers; the dashed line is chance. The CNN and the landmark classifiers, but not the centroid-size classifier, score in the high 0.80s on the full set and near chance on the within-study half. Values for all classifiers, with intervals, are in \siref{MS-SI-WB-BASELINES}, and per-stratum Stage 2 baseline accuracies are in \siref{MS-SI-WB-STAGE2-PREDICTION-PROVENANCE}.}
\label{fig:MS-FIG-POOLED-DESIGN}
\end{figure}

To test whether this concentration of predictions within the specimen's study is specific to deep learning, we evaluated three baseline classifiers on the same training and test partitions (\cref{fig:MS-FIG-POOLED-DESIGN}; \methodsref{sec:baseline-classifiers}). These comprised the study-only classifier, which sees only each specimen's study; the centroid-size classifier, standardized within stratum so that it cannot identify the study; and the landmark classifier, a standard geometric-morphometric method.\cite{zelditch2012}

The study-only classifier and the Stage 1 CNN exhibit the same split, achieving near-perfect accuracy on the strain panels and mutant-only studies while dropping to or near chance on the within-study subset (\cref{fig:MS-FIG-POOLED-DESIGN}C). Next, the study-blind centroid-size classifier performs near chance at Stage 1 and poorly at Stage 2 (\crefpanels{fig:MS-FIG-POOLED-DESIGN}{A}{C}). The landmark classifier based on 94 sparse cranial landmarks, by contrast, mirrors the CNNs. At Stage 1, it reaches 0.829 balanced accuracy on the full test set but falls near chance on the within-study subset. At Stage 2, it identifies the study far more reliably than the specific gene label, with most errors remaining inside the specimen's true study (\cref{fig:MS-FIG-POOLED-DESIGN}A--C). The landmark classifier fit on all 844 cranial points shows the same pattern (\siref{MS-SI-WB-BASELINES}). Because classifiers based on landmark shape behave like the CNNs, these results indicate that the shortcut structure is inherent to this pooled design rather than an artifact of deep learning.

Beyond this shared structure, the Stage 1 CNN retains a performance margin over all baselines on the within-study subset (AUC 0.729 [0.646, 0.807] versus 0.610 [0.521, 0.695] for sparse landmarks; \siref{MS-SI-WB-BASELINES}).Whether this modest advantage arises within studies and reflects morphology is examined in the matched cohort below.

\subsection{Modest, imprecise signal remains within study and strain}\label{sec:residual-signal}

To estimate how much signal remains once study and strain are held fixed, we then evaluated specimens only within the same study and strain. This performance cannot be attributed to the target phenotype alone. Sex, zygosity, treatment, and exact age were not balanced within study and strain, and litter and batch were not recorded. Within each multi-label study and within each background strain in the Enhancer study, the network outperforms the majority-class baseline across both partitions, though confidence intervals are wide and no margin exists for \emph{Isl1} versus \emph{Msx1} (\cref{fig:MS-FIG-STAGE2-ERRORS}C; \siref{MS-SI-WB-STAGE2-PREDICTION-PROVENANCE}). For control versus mutant, among the three studies with more than five within-study controls, within-study AUC is highest for Spry (0.85) and Longshanks (0.77) and near chance for Enhancer (0.59) (\siref{MS-SI-WB-STAGE1-PREDICTION-PROVENANCE}). The Longshanks cohort provides the cleanest internal comparison because its mutants are selected-line CD-1 animals from a single study and its controls share the same outbred background. Together, these comparisons indicate a genuine but modest and imprecise signal within studies.

\subsection{Matching on recorded provenance yields modest, study-specific discrimination}\label{sec:matched-cohort}

Because the within-study estimates above still derive from models trained on the confounded pool, we next measured residual morphological association in a matched cohort where the matched variables (study, strain, life phase, and sex) cannot predict the label (\methodsref{sec:matched-cohort-and-grouped-splits}). Designed as a controlled follow-up on the same resource rather than an external replication, this cohort excluded the strain panels, mutant-only studies, Longshanks lines, and treated animals. We balanced controls and mutants within strata defined by shared study, strain, and life phase (with exact sex balance). The resulting cohort comprises 924 specimens across 16 strata from seven studies (\cref{fig:MS-SI-FIG-COHORT-FUNNEL}; \siref{MS-SI-WB-MATCHED-COHORT}). A metadata-only classifier evaluated on this cohort performs exactly at chance, confirming that recorded matching variables no longer predict condition.

\begin{figure}[htbp]
\centering
\includegraphics[width=\textwidth]{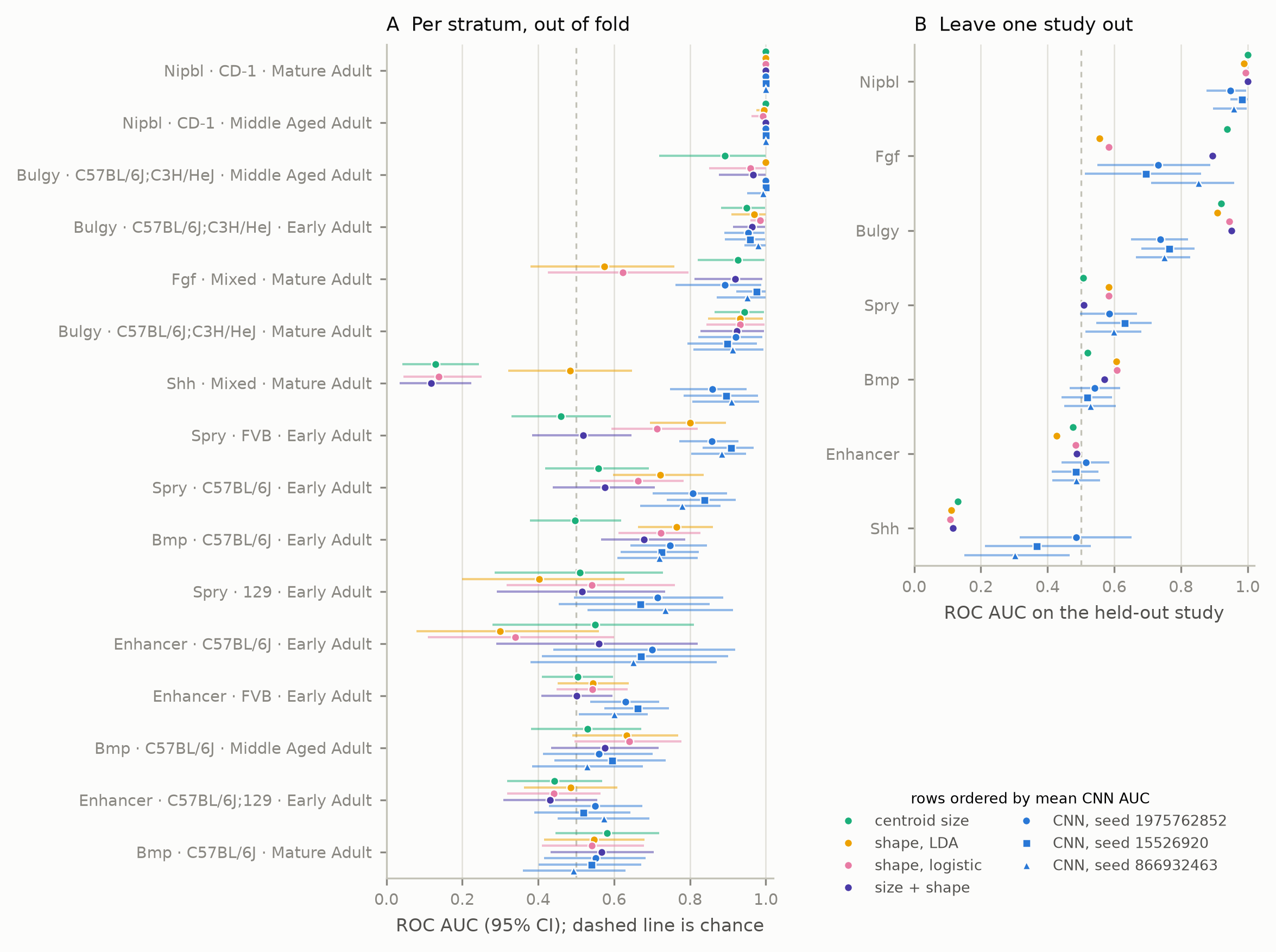}
\caption{With recorded provenance matched, discrimination is modest, study-specific, and transfers inconsistently across studies. The control-versus-mutant CNN was retrained on the matched cohort and is shown beside the landmark and centroid-size classifiers on the same splits and specimens. (A) Out-of-fold ROC-AUC within each study $\times$ strain $\times$ life-phase stratum, with class-stratified bootstrap intervals, for the CNN at each of three training seeds and the four baselines once each. Rows are ordered by mean CNN AUC, and the dashed line is chance. The CNN's point estimates exceed the baselines where size provides little information and match them where size is decisive. In Shh, the CNN separates the groups even though the centroid-size classifier scores controls above mutants; this comparison is descriptive, with no paired interval. (B) AUC on each held-out study in leave-one-study-out training. The CNN transfers no better than the baselines, and its AUC falls below 0.5 for Shh.}
\label{fig:MS-FIG-MATCHED-CNN}
\end{figure}

On this cohort, we retrained the centroid-size classifier, landmark classifiers, and control-versus-mutant CNN (across three random seeds using its original configuration) via repeated 5-fold cross-validation and leave-one-study-out evaluation (\cref{fig:MS-FIG-MATCHED-CNN}; \sirefs{MS-SI-WB-BASELINES}{MS-SI-WB-MATCHED-COHORT}). In cross-validation, the landmark and centroid-size classifiers achieve a stratum-averaged ROC-AUC of 0.66--0.72, but this average masks substantial variation across strata. Discrimination is near-perfect in the Bulgy (\emph{Fgfr2}), Nipbl, and Fgf strata, where mutants differ markedly from controls in overall cranial size, and substantially lower elsewhere. The centroid-size classifier falls far below chance in Shh, where mutants are distinctly larger than controls whereas elsewhere they are smaller or similar in size, because it fits a single direction of size effect across all strata (\siref{MS-SI-WB-MATCHED-COHORT}).

We next compared the retrained CNN against these baselines within individual strata to assess whether deep learning captures non-size morphological features. At each random seed, the CNN achieves a higher stratum-averaged AUC (0.77--0.79) than the best landmark classifier (0.72). However, a paired comparison across identical strata and folds yields a CNN$-$landmark AUC difference of 0.047 [$-$0.008, 0.120] at the first seed, with confidence intervals spanning zero at two of the three seeds (\methodsref{sec:paired-comparison-on-the-matched}; \siref{MS-SI-WB-MATCHED-COHORT}). Because strata are nested within seven studies, these intervals describe the sampled strata rather than a broader population of studies, rendering the comparison descriptive. The CNN's apparent advantage appears mainly in the Spry and Enhancer strata, where global size is uninformative, and in Shh, where it separates mutants from controls despite the inverted size effect. It disappears where the centroid-size and landmark classifiers are already near-perfect. These comparisons show that matching changes how every classifier's performance is read, without establishing that any one classifier is superior.

Finally, we tested transfer across studies by holding out one study at a time during training (\cref{fig:MS-FIG-MATCHED-CNN}B). Across held-out studies, the CNN transfers no better than the baselines, yielding a mean leave-one-study-out AUC of 0.63--0.65 compared to 0.54--0.65 for the baselines. Every classifier performs well on certain held-out studies and near chance on others. Shh shows the contrast directly: the CNN separates its mutants from controls when Shh specimens are included in training, but when Shh is held out its AUC drops to or below chance alongside every baseline, as if smaller cranial size were a transferable signature of mutation. Moreover, the network fits its training data nearly perfectly and remains as overconfident as the original Stage 1 CNN (\siref{MS-SI-WB-MATCHED-COHORT}). Together, these results provide no evidence of a consistent control-versus-mutant morphological signature that transfers across the studies and classifiers evaluated.

%% file: sections/04_discussion.tex
\section{Discussion}\label{sec:discussion}

\subsection{Apparent accuracy of models trained on pooled collections can reflect provenance}\label{sec:disc-confounding}

Our results show that high classification accuracy and biologically plausible error patterns in pooled biological datasets can arise from source-nested label confounding rather than the target phenotype. This confounding is a byproduct of standard study design, not an oversight by individual contributors. A study typically investigates one condition or a few related ones, compares them with its own controls, and often uses a distinct biological background and acquisition setup, so pooling such studies makes each condition collinear with its source and much of its provenance. Here, each of the 16 gene labels came from a single study, even the broad control-versus-mutant label was strongly associated with study, and the gene-label network's predictions and errors stayed almost entirely within each specimen's own study and background strain. A landmark classifier showed a similar pattern, which points to the pooled design rather than the classification method.

More broadly, when labels are nested within sources, a model that learned the target phenotype would produce the same pattern of correct predictions and errors as one that recognizes sources. Standard evaluation metrics therefore cannot separate biological signal from source structure. To address this gap, we provide six diagnostic checks (\cref{tab:MS-TAB-DIAGNOSTICS}) for recognizing source-nested label confounding before any biological interpretation is attempted. For researchers developing or using classifiers trained on pooled resources, the practical implication is that high accuracy and biologically plausible mistakes are not evidence that a model has learned the phenotype of interest until such checks have been run.

\subsection{Why conventional evaluation was misleading}\label{sec:why-conventional-evaluation-was-misleading}

Our evaluation used safeguards common in sample-level analyses: disjoint sample identifiers, bootstrap intervals, permutation tests, per-class exact intervals, calibration assessment, and withholding of low-confidence predictions. Each guards against a genuine pitfall, such as the same sample appearing in training and test, sampling noise, or error patterns arising by chance, but none tests whether the label is collinear with provenance. A random sample-level split ensures that no test sample was seen in training, but other samples from the same source and background (here, study and strain), and possibly the same acquisition batch, were typically in training with the same label. As a result, a classifier can score well simply by recognizing where a sample came from.

The error patterns made this failure convincing. Nearly all errors confused labels that share a study and background strain, and for many of them the contributing studies' own publications supply a biological explanation. Confusing one Sprouty gene with another,\cite{percival2017} for example, is what a classifier sensitive to morphology might be expected to do. But sources typically study related conditions together, so a classifier that tracks provenance makes the same biologically coherent errors. A plausible explanation for an error is therefore expected under source-nested label confounding and is not evidence that the model learned the phenotype.

\subsection{What residual biological signal can and cannot be inferred}\label{sec:what-residual-biological-signal-can}

Of the types of provenance distinguished in the \secref{sec:introduction}, the confounding we could measure involves study design and biological background provenance. The control-versus-mutant classifier tracks background strain, a biological variable but not the one the label denotes, and technical acquisition provenance cannot be assessed because it is not recorded. Background strain is also more than a confound, because in the Spry study the size and direction of the mutations' effects on skull shape depended on it.\cite{percival2017} What remains once recorded provenance is held fixed is modest, imprecise, and not attributable to target morphology alone. In the matched cohort, where the matched variables cannot predict the label, landmark and centroid-size classifiers separate controls from mutants mainly in studies whose mutants differ in overall skull size.

A CNN retrained on the matched cohort does better than these classifiers, mostly where size is uninformative, although its advantage over the best landmark classifier is uncertain. That margin could reflect morphology that landmarks miss or scanning differences within a study, which image intensities record but landmark coordinates do not. Testing on held-out studies does not settle the question. Discrimination proved study-specific when whole studies were held out. This is expected even if the signal is morphological, because mutations in different genes can produce distinct skull phenotypes. Failure to transfer is therefore consistent with the signal reflecting either morphology or scanning differences. These data contain real within-study discriminative information, but attributing it to morphology requires recorded scan session and batch.

\subsection{Diagnostics for source-nested label confounding}\label{sec:practical-diagnostics-and-implications-for}

The provenance audit in the \secref{sec:results} (\emph{\nameref{sec:provenance-audit}}) forms six reusable diagnostics (\cref{tab:MS-TAB-DIAGNOSTICS}), each requiring only recorded provenance and the predictions a standard evaluation already produces. Here, each sample's source is its contributing study and its background is mouse strain. In other pooled collections, the source may be a recruiting site, hospital, or biobank, and the background may be ancestry, cell line, or donor age. Together the diagnostics assess whether the label is predictable from provenance (1--2), whether the classifier's behavior is separable from provenance (3--4), and whether claims about its errors survive conditioning on recorded provenance (5--6). The first three apply to any supervised prediction, such as regression of a continuous trait, with the target in place of the label and the corresponding error measure in place of accuracy, whereas the last three apply only to classification, because they rely on categorical errors.

None of the six is new in isolation. Tabulating label against source and fitting metadata-only baselines (1--2) is standard confounder analysis,\cite{badgeley2019,wachinger2021} while evaluating within sources (3) follows the blocked cross-validation literature.\cite{roberts2017cv,saeb2017} Similarly, measuring the probability assigned to labels from the specimen's own source (4) resembles testing whether a medical-imaging model can identify which hospital an image came from.\cite{zech2018} Finally, comparing errors with the labels each source contributes (5) is a simple tabulation, and requiring per-class effects to hold in both validation and test (6) acts as a leakage-aware replication check.\cite{kaufman2012,kapoor2023} Existing tests detect whether a model has learned a confounder by shuffling labels within a source\cite{chaibubneto2019} or by resampling the source given the label.\cite{spisak2022} Both presuppose overlap. Where a source contributes one label, or a label comes from one source, there is nothing to shuffle or resample, which is why the audit begins by tabulating labels against sources. Our contribution is assembling these checks into a stepwise audit for resources where labels and sources are collinear by design and holding out sources is impossible for some labels.

\begin{table}[htbp]
\centering
\footnotesize
\caption{Six diagnostics for source-nested label confounding in classifiers trained on pooled biological data.}
\label{tab:MS-TAB-DIAGNOSTICS}
\begin{tabular}{p{0.2cm}@{\hspace{5pt}}>{\raggedright\arraybackslash}p{3.5cm}>{\raggedright\arraybackslash}p{3.4cm}>{\raggedright\arraybackslash}p{3.0cm}>{\raggedright\arraybackslash}p{4.54cm}}
\toprule
& \textbf{Diagnostic} & \textbf{Question; permits} & \textbf{Cannot establish} & \textbf{Result here (test set)} \\
\midrule
1 & \textbf{Label--source nesting.} Cross-tabulation of label by source and other recorded provenance (e.g., strain or ancestry, sex, age, scanner), showing how far each label is confined to one source and how well source predicts it. & Can the label be predicted from the sample's source alone? Permits identifying the finest level (source, or finer strata of recorded provenance) at which claims must be evaluated. & Whether the model uses the nesting; which type of provenance (acquisition, study design, or biological background) underlies it. & Gene label: H(Study $\mid$ Label) = 0; H(Label $\mid$ Study) = 1.21 of 3.86 bits; 6 of 16 labels alone in their study; zygosity and sex confounded with the label in one Enhancer stratum. Control versus mutant: H(Group $\mid$ Study) = 0.41 of 1.00 bits. \\
\addlinespace[2pt]
2 & \textbf{Provenance-only baseline.} Performance of a classifier given only recorded provenance, and of a majority-class baseline within each provenance group. & How much of the classifier's score can recorded provenance alone reach? Permits setting the reference against which pooled accuracy is read. & How much of the model's accuracy provenance causes; in a pooled split, provenance and phenotype predict the same labels, so their contributions cannot be separated. & Study-only classifier 0.874 for control versus mutant (CNN 0.872); six-field metadata-only classifier 0.720 for gene label (CNN 0.848); majority-class baseline 0.712 within study and strain. \\
\addlinespace[2pt]
3 & \textbf{Within-source evaluation.} Performance within each source (and finer strata of recorded provenance) and on sources holding more than one label, against majority-class, provenance-only, and conventional baselines. & Does discrimination survive when samples are compared only with recorded provenance held fixed? Permits an estimate of the discrimination that recorded provenance does not explain. & Whether surviving discrimination reflects the target phenotype, remaining background or design differences, or acquisition differences. & Control versus mutant: specificity on within-study controls 0.32; AUC on the within-study subset (nine studies pooled) 0.73 vs 0.61 for cranial landmarks, and 0.59--0.85 within the three studies with more than five test controls. Gene label: accuracy within study and strain 0.50--0.85 vs majority-class baseline 0.48--0.54. \\
\addlinespace[2pt]
4 & \textbf{Own-source concentration.} Probability assigned to labels from the sample's own source, and the share of errors that cross sources, before and after calibration. & Do mistakes stay within the sample's own source? Permits identifying cross-source errors, where provenance and target-phenotype explanations differ. & That the model recognizes the source; a classifier that identifies the target phenotype would also stay within its source. & Own-study probability 0.988 (0.891 calibrated); 2 of 20 errors cross studies; accuracy is unchanged when predictions are restricted to the true study's labels. \\
\addlinespace[2pt]
5 & \textbf{Error-pattern check.} Comparison of where the errors fall with the labels available in each sample's own source, a grouping that study design fixes independently of the predictions. & Does the pattern of mistakes (e.g., within gene families or disease subtypes) add evidence beyond recorded provenance? Permits retaining or dropping the pattern as incremental evidence. & Whether errors within a source reflect morphology or provenance. & 18 of 20 errors fall on the 8 label pairs (of 120) that share a study and strain, against 2.2 expected at random; the contributing studies document biological relationships for two of the four groups. \\
\addlinespace[2pt]
6 & \textbf{Cross-partition consistency.} Per-class asymmetries compared between test and validation, interpreted only where they agree. & Does a per-class pattern hold in both test and validation? Permits declining to interpret asymmetries that fail. & That the pattern would replicate in independent data, because test and validation come from the same sources. & The \emph{Spry4} absorption and the \emph{Hdac9} $\rightarrow$ \emph{Twist1} bias reverse between test and validation; \emph{Isl1} and \emph{Msx1} are exchanged at chance in both. \\
\bottomrule
\end{tabular}
\end{table}

\subsection{Implications for model evaluation and data curation}\label{sec:disc-implications}

The main implication for evaluation is to match the evaluation design to the claim. To claim that a classifier will perform well on new samples from the same contributors, a random sample-level split is appropriate, because it estimates performance on samples whose provenance is already represented in training. To claim that a classifier detects label-associated differences, the evaluation must hold provenance fixed, comparing samples only within the same source and, where recorded, the same biological background and acquisition batch. Labels recorded in only one source remain usable for this claim whenever their source also contributes other labels, as Spry and Enhancer do here, or its own controls. To claim that a classifier transfers to new sources, leave-one-source-out evaluation is needed. It can assess labels recorded in several sources, as controls and mutants are here, but not a label recorded in only one source, which has no training examples once that source is held out (here, all 16 gene labels). Under source-nested label confounding, performance on a random split can greatly overstate how well a classifier detects label-associated differences or transfers to new sources, so it should not be presented as evidence for either claim.

Running these diagnostics requires recording provenance alongside the label: at minimum each sample's contributing source, together with its biological background (genetic background such as strain or ancestry, sex, age, and zygosity) and any treatment that varies within a source, since each of these can confound comparisons made within that source. Relatedness, such as litter or family structure, and acquisition batch are needed to bound effects within a source. Where these fields are missing, the diagnostics cannot be fully run, the absence should be reported as a limitation, and claims should be restricted to what a random split measures. This hazard does not mean that pooled resources are flawed. Much of the variation between sources and backgrounds is the biology the contributing studies set out to measure. The problem arises only when a random sample-level split is treated as an independent test of biological generalization. Two practices would reduce the hazard. The first is to distribute a summary table alongside the data, cross-tabulating every recorded variable by source, background, and acquisition, so that users can see how strongly source predicts any variable they might choose as a label before model development. The second is for each source to deposit controls alongside its cases in comparable numbers, with relatedness and acquisition batch recorded.

\subsection{Limitations of the study}\label{sec:limitations-of-the-study}

The diagnostics, and our demonstration of them, have limits of their own. We demonstrated them on one pooled resource and one imaging modality, with two classification tasks and two kinds of image-based classifier (deep learning and landmark morphometrics). How strongly source-nested label confounding affects another collection depends on how tightly its labels are nested within sources, which the label--source nesting check (Diagnostic 1) measures. Our exemplar is also an extreme case, with every gene label confined to a single study. It shows what the diagnostics reveal under strong nesting, but not how sensitive they are when nesting is partial or confounding is weak. Measuring that sensitivity would require pooled collections with partial nesting. The diagnostics can detect nesting only in recorded provenance, not in unrecorded factors such as litter or acquisition batch. They also quantify confounding without setting a threshold for when it is too strong to ignore. The error-pattern check (Diagnostic 5) uses a grouping set by the contributing studies, namely which labels each recorded on each background strain, so it does not depend on the predictions. It cannot show whether errors within a source reflect morphology or provenance, and any finer grouping of related labels would have to be fixed before predictions are examined, which we did not do. Finally, agreement in the cross-partition consistency check (Diagnostic 6) is not a replication, because test and validation come from the same sources.

Gaps in recorded provenance limit the estimates of residual signal, not the findings about the pooled design. Those findings rest on the association of label with study and strain, a property of the metadata, and are reproduced by landmark classifiers. The splits were grouped only by biosample identifier, so repeat scans, relatives, or shared acquisition batches may cross partitions. They were stratified by sex and life phase across the pooled data rather than within each study, so residual performance within study and strain may reflect covariates still confounded with the label (sex, zygosity, treatment, exact age). Even the matched cohort cannot balance exact age, litter, or scan session. Technical acquisition differences are the factor most likely to affect the CNN but not the landmark classifiers, which do not use image intensities. The specimens were acquired on three scanners without intensity harmonization (\methodsref{sec:data-source-and-specimen-selection}). Because the scanner for each specimen is not recorded, no image-level correction can separate scanner or session from control or mutant status where they coincide within a study.

Precision is also limited. Several estimates within study and strain rest on 10--24 specimens, and several matched-cohort strata are at the minimum of five specimens per group. The original control-versus-mutant and gene-label CNNs were each trained once, at one seed and on volumes downsampled to 48 $\times$ 48 $\times$ 48 voxels, so the residual signal and the control-versus-mutant CNN's margin over the landmark classifiers are specific to this configuration. The individual errors of the gene-label CNN, and therefore the count of errors within study and strain and the per-class asymmetries, likewise describe one trained model, and another seed would produce a different set of errors. Variation across training seeds was assessed only in the matched cohort, where the interval for the CNN's margin over the best landmark classifier included zero at two of three seeds. Both methods may also understate what is detectable: the CNN because it used volumes downsampled without anti-aliasing and rescaled to each volume's maximum intensity, and the landmark classifiers because they used a fixed set of landmarks. A network with higher resolution or better intensity normalization might therefore detect more signal within study and strain. That would not alter the findings about the pooled design, which do not depend on the network, and any gain could be attributed to morphology only with recorded scan session and batch.

%% file: sections/05_methods.tex
\section{Methods}\label{sec:methods}

\subsection{Data source and specimen selection}\label{sec:data-source-and-specimen-selection}

All skull volumes and landmark coordinates were obtained from MusMorph,\cite{devine2022} a public database of standardized mouse morphology distributed through FaceBase\cite{schuler2022facebase,samuels2020facebase} under project identifier FACEBASE:3-HXMC.\cite{musmorph_project} On FaceBase, MusMorph is organized as one dataset per contributing study, plus a project-wide dataset that holds the per-specimen metadata file (\texttt{MusMorph\_\allowbreak Metadata.csv}) and the study metadata file (\texttt{Study\_Metadata.csv}).\cite{musmorph_projectwide} Our analysis uses the adult skull scans (\texttt{Anatomy} = \texttt{Head;Skull}), which we call the \emph{adult skull pool} (5,814 specimens in the metadata retrieved on 19 November 2025). These specimens come from 23 of the 28 study datasets,\cite{musmorph_b9d,musmorph_bmp,musmorph_brachymorph,musmorph_brainface,musmorph_bulgy,musmorph_collaborativecross,musmorph_diversityoutbred,musmorph_enhancer,musmorph_fgf,musmorph_ghrhr,musmorph_hybrid,musmorph_ift,musmorph_longshanks,musmorph_lrp,musmorph_mps,musmorph_nipbl,musmorph_nosip,musmorph_osteoimperfecta,musmorph_pten,musmorph_shh,musmorph_spry,musmorph_straincomparison,musmorph_trp} whose identifiers are listed in the key resources table. The other five (Ap2, IMPC, Mks, Tctn, and Wnt) contain only embryonic scans. The specimens are postnatal \emph{Mus musculus} of both sexes, aged P14 to P533, which MusMorph's \texttt{Life\_Phase} field groups into Early, Mature, and Middle-Aged Adult (Early Adult spans P14 to P90 and so includes juveniles). They come from inbred, outbred, hybrid, and mutant lines on the genetic backgrounds recorded in MusMorph's \texttt{Strain} field, including C57BL/6J, C57BL/6J;129, 129, FVB, CD-1, C57BL/6J;C3H/HeJ, and mixed backgrounds (\cref{fig:MS-FIG-PROVENANCE-STRUCTURE}; \siref{MS-SI-WB-STAGE2-SPLIT-PROVENANCE}). We performed no additional scanning or animal work. Ethical approval, husbandry, and acquisition details for each contributing line are documented in MusMorph's per-dataset records on FaceBase and in the primary publications listed for each study in the study metadata file.

The MusMorph authors report that these specimens were scanned unfixed and unstained under a standardized protocol at the University of Calgary on three micro-CT scanners at voxel resolutions of 0.035--0.044 mm.\cite{devine2022} Scanner, acquisition session, operator, and fixation batch are not among the per-specimen metadata fields, so technical acquisition provenance is not observable in this study. MusMorph supplies each adult specimen as a volume, after format conversion, resampling to the adult atlas dimensions (642 $\times$ 586 $\times$ 979 voxels), and rigid alignment (translation and rotation only) to the adult cranial atlas. The MusMorph authors list intensity correction and intensity normalization among the preprocessing steps for embryo images only,\cite{devine2022} so the adult volumes have no harmonization of intensity across scanners or acquisition settings, and our pipeline adds none beyond the per-specimen rescaling described in \emph{\nameref{sec:preprocessing}}. MusMorph propagated landmarks to each specimen by registration to a pre-existing wild-type adult atlas. MusMorph does not identify the atlas specimens, so they may include specimens in our partitions.

\subsection{Classification targets and provenance fields}\label{sec:classification-targets-and-provenance-fields}

The classification targets and provenance fields are columns of the per-specimen metadata file (\texttt{MusMorph\_\allowbreak Metadata.csv}). The training targets are \texttt{Experimental\_\allowbreak Group} (\texttt{Control} or \texttt{Mutant}) for the Stage 1 control-versus-mutant classifier and the \texttt{Gene} field for the Stage 2 gene-label classifier, with \texttt{Biosample} identifying each volume. We call the \texttt{Gene} field the gene label. It is the Mouse Genome Informatics (MGI) symbol where one exists. Two labels are preserved as MusMorph records them: \emph{Kcna1;Crf4} (\emph{Crf4} unspecified in the \texttt{Genotype} field) and \emph{hs1629} (the VISTA Enhancer Browser \cite{visel2007} convention for a conserved non-coding element). A gene label therefore does not always name a single gene. We preserve every label and genotype string as recorded and treat the 16 gene labels as recorded labels rather than verified allele designations. A label does not indicate the kind of allele, for example whether it alters a gene's coding sequence or a nearby regulatory element, and we did not verify the allele behind each label. \texttt{Experimental\_\allowbreak Group} = \texttt{Mutant} also includes the Longshanks selected-line animals (\texttt{Zygosity} = \texttt{Wildtype}, no \texttt{Gene}), as MusMorph documents for that study. The network's only input is the image volume. \texttt{Anatomy}, \texttt{Life\_Phase}, \texttt{Sex}, and \texttt{Availability} define the cohort and the split strata, and \texttt{Dataset} (the contributing study), \texttt{Strain}, \texttt{Strain\_Type}, \texttt{Zygosity}, \texttt{Genotype}, and \texttt{Treatment} are provenance fields used in the analysis. MusMorph does not record litter, so litter structure is not observable in this study.

\subsection{Splits and dependence control}\label{sec:splits-and-dependence-control}

Two split tables were derived from MusMorph metadata. Both come from one 70/15/15 assignment of the released specimens that have recorded sex and life phase, stratified by life phase and sex and, for mutants, by gene label. The Stage 1 table (\texttt{mutant\_vs\_control\_\allowbreak balanced.csv}) contains 3,124 specimens (training 2,202 [1,101 controls + 1,101 mutants]; validation 500 [250 + 250]; test 422 [211 + 211]), with controls subsampled at random within each partition and life phase to the number of mutants, so that every eligible mutant was retained. The Stage 2 table (\texttt{gene\_classification.csv}) contains 908 mutant specimens (training 633; validation 143; test 132) from the 16 labels with at least 30 mutant specimens in the cohort, counted across all partitions before splitting. Sex was approximately balanced overall, with 1,575 females and 1,549 males at Stage 1 and 424 females and 484 males at Stage 2. Both tables include specimens with a recorded treatment value (503 at Stage 1, 162 at Stage 2).

No biosample appears in more than one partition of either table, all 132 Stage 2 test specimens are Stage 1 test mutants, and no test specimen of either stage appears in the other stage's training or validation partition. This is the only dependence that was controlled. The splits were not grouped by litter, breeding cohort, scan session, or acquisition batch, which MusMorph does not record, and neither split was stratified by study or strain.

\subsection{Preprocessing}\label{sec:preprocessing}

Each MusMorph volume is loaded with NiBabel, resampled from the 642 $\times$ 586 $\times$ 979 atlas array to 48 $\times$ 48 $\times$ 48 with \texttt{scipy.ndimage.zoom} (linear interpolation), and divided by its own maximum intensity, so that each volume's maximum is 1. Because each axis is resampled independently, the working voxel is anisotropic (approximately 0.468 $\times$ 0.427 $\times$ 0.714 mm; dimensions here and above are in world x, y, z order) and the skull is rescaled into a cube. No low-pass (anti-aliasing) filter was applied before downsampling, and field of view and padding follow the atlas rather than the individual skull.

\subsection{Network and training}\label{sec:network-training}

Both stages use a 3D ResNet-18 backbone implemented in PyTorch (\cref{fig:MS-FIG-ARCHITECTURE}). It is the standard ResNet-18\cite{he2016} with kernels extended from 2D to 3D (33.2 M parameters), with dropout (p = 0.5) before the 16-logit Stage 2 head. Each stage was trained from scratch in a single run on one GPU without mixed precision, at random seed 1191077522, drawn once from the operating-system random source and recorded in the configuration. Training used Adam (learning rate 10$^{-4}$, weight decay 10$^{-4}$), class-weighted cross-entropy, gradient clipping at 1.0, and ReduceLROnPlateau (factor 0.5; patience 3 at Stage 1 and 5 at Stage 2). Stage 1 used batch size 32 for up to 30 epochs with early stopping on validation balanced accuracy (patience 10), and Stage 2 used batch size 4 for up to 75 epochs with early stopping on validation macro-average recall (patience 20). The checkpoint from the epoch with the best value of the early-stopping metric was retained. No data augmentation was used. The layer specification, per-epoch training curves, and configuration and checkpoint hashes are in the code release.

\subsection{Evaluation protocol}\label{sec:evaluation-protocol}

Each test set was evaluated once, after all architecture and training decisions had been fixed on the validation set. Stage 1 primary metrics are at the default 0.50 threshold. The threshold sweep and the Youden- and F1-optimal thresholds are post hoc characterizations (\cref{fig:MS-SI-FIG-THRESHOLD-SWEEP}; \siref{MS-SI-WB-STAGE1-PREDICTION-PROVENANCE}). Stage 2 metrics that use hard predictions take the argmax of the 16-class softmax, and probability-based metrics use the full distribution. Joint performance of the two stages on the 132 shared test specimens, counting a specimen as correct only if Stage 1 calls it mutant and Stage 2 assigns its gene label correctly, is reported in \siref{MS-SI-TAB-CONDITIONAL-ATTRIBUTION}.

\subsection{Study and strain decomposition}\label{sec:study-and-strain-decomposition}

The split tables were joined to the MusMorph metadata table on \texttt{Biosample}. For Stage 2 we tabulated Label $\times$ Study and Label $\times$ Split by \texttt{Strain}, \texttt{Zygosity}, \texttt{Life\_Phase}, and \texttt{Sex} (Diagnostic 1; \siref{MS-SI-WB-STAGE2-SPLIT-PROVENANCE}). A specimen's Stage 2 stratum is its contributing study, with the Enhancer study divided by background strain, because each Enhancer label was recorded on a single background across all partitions whereas each Spry label was recorded on all three of that study's backgrounds. The labels in a specimen's stratum are therefore exactly the labels recorded in its study on its background strain, so the stratum is the finest level at which the gene labels are source-nested (\secref{sec:introduction}). Entropies are in bits, H(X $\mid$ Y) is the Shannon entropy of X remaining once Y is known, and H(Label $\mid$ Study) on the test set is the mean of within-study label entropies, with each study weighted by its number of test specimens.

Two baselines are given the true stratum of every test specimen, information the CNN never receives. The \emph{uniform-guess baseline} scores the expected accuracy of guessing uniformly among the training labels in each specimen's stratum. The \emph{majority-class baseline} predicts, for every specimen in a stratum, the label with the most training specimens in it (Diagnostic 2). Its accuracy within a stratum, the proportion of evaluation specimens that have that label, is the reference for the within-study and within-stratum accuracies (Diagnostic 3). From the per-specimen predictions (\siref{MS-SI-WB-STAGE2-PREDICTION-PROVENANCE}), we computed for Diagnostic 4 the study of the predicted label, the own-study probability (the softmax probability assigned to labels of the true study, before and after temperature scaling), and the study-restricted prediction (the argmax over the true study's labels only).

For Stage 1 we tabulated Study $\times$ Split $\times$ \texttt{Experimental\_\allowbreak Group} and \texttt{Strain} $\times$ (\texttt{Experimental\_\allowbreak Group}, Study), and computed H(Group $\mid$ Study) on the training partition (Diagnostic 1; \siref{MS-SI-WB-STAGE1-SPLIT-PROVENANCE}). Studies with test specimens were classified as control-only (the four strain-panel studies CC, DO, Hybrid, and Strain\_Comparison), mutant-only, or containing both controls and mutants. Test controls from these two kinds of study are the strain-panel and within-study controls defined in the \secref{sec:results}. The within-study subset comprises all test specimens from the nine studies that contribute both test controls and test mutants (Bmp, Bulgy, Enhancer, Fgf, Ift, Longshanks, Nipbl, Shh, Spry). Specificity, sensitivity, balanced accuracy, and ROC-AUC were computed on this subset, its complement, the full test set, and per study within the subset (Diagnostic 3; \siref{MS-SI-WB-STAGE1-PREDICTION-PROVENANCE}).

\subsection{Baseline classifiers}\label{sec:baseline-classifiers}

Three non-CNN classifiers were fit on the training partition of each existing split and evaluated on its validation and test partitions. The metadata-only classifier is the provenance-only baseline of Diagnostic 2, and the centroid-size and landmark classifiers test whether the patterns seen with the CNN depend on the classification method. The centroid-size and landmark classifiers use the MusMorph per-specimen landmark files (coordinates in native specimen space). Every cranial file contains 94 sparse landmarks, although the MusMorph descriptor states 93,\cite{devine2022} and we use all 94. Thirteen adult specimens have no landmark files and are excluded from landmark and centroid-size results (two Stage 1 and one Stage 2 test specimens; listed per analysis in \siref{MS-SI-WB-BASELINES}). No landmark-error screen was applied, and the largest Procrustes distances, in a few specimens, are about seven times the median.

(i) \emph{Metadata-only classifier.} An L2-regularized (multinomial) logistic regression on one-hot \texttt{Dataset}, \texttt{Strain}, \texttt{Sex}, \texttt{Life\_Phase}, \texttt{Zygosity}, and \texttt{Treatment}, with the inverse regularization strength $C$ chosen by validation accuracy from \{0.01, 0.1, 1, 10, 100\}, is reported for \texttt{Dataset} alone (the study-only classifier) and for all six fields. Since \texttt{Zygosity} encodes the Stage 1 label by definition, the Stage 1 comparison uses the study-only classifier. 

(ii) \emph{Centroid-size classifier.} Centroid size is the root summed squared distance of the 94 sparse cranial landmarks from their centroid, computed from raw coordinates before superimposition. It was standardized within stratum with the training-partition mean and standard deviation, which removes differences in mean and scale between strata, and entered into an L2 logistic regression. The stratum is the (\texttt{Dataset}, \texttt{Strain}) pair on the existing splits and the cohort stratum on the matched cohort, and a stratum with fewer than two training specimens uses the pooled training values.

(iii) \emph{Landmark classifier.} The 94 sparse cranial landmarks were superimposed by generalized Procrustes analysis,\cite{zelditch2012} with the Procrustes mean estimated on the training partition, and validation and test specimens were aligned to that fixed mean in one step. Superimposition was partial Procrustes (unit centroid size, rotation without reflection) without full-Procrustes rescaling or tangent projection. Principal component analysis (PCA) was computed on the training Procrustes coordinates. A shrinkage (Ledoit--Wolf) linear discriminant served as the primary classifier and L2 multinomial logistic regression as the secondary. Both used the leading principal components, with the number chosen by validation accuracy from \{2, 3, 5, 8, 10, 15, 20, 30\} and capped at $\max(2, \lfloor n_{\mathrm{train}}/10 \rfloor)$, and ties resolved in favor of fewer components or smaller $C$ (stronger regularization). At Stage 2 the landmark classifier was fit both on all training specimens and separately inside each evaluation stratum.

Secondary runs used all 844 cranial points (the 94 landmarks plus 750 curve and surface semilandmarks, treated as fixed points rather than slid,\cite{gunz2013} because MusMorph's registration sets their correspondence) and the 19 mandibular and 12 endocast sparse landmarks as separate structures. They also used combined models with a separate superimposition and PCA for each structure and, on the matched cohort only, a combined size-and-shape logistic model. Component-selection details for every combination are in \siref{MS-SI-WB-BASELINES}. CNN reference values beside the baselines are computed on the full test sets. Intervals are Clopper--Pearson for proportions and class-stratified bootstrap (10,000 draws) for balanced accuracy, AUC, and own-study probability.

\subsection{Matched cohort and grouped splits}\label{sec:matched-cohort-and-grouped-splits}

The matched cohort is designed so that the matched variables (study, strain, life phase, and sex) cannot predict the label. It was built from the adult skull pool of controls and mutants by sequential exclusion steps, with counts logged at each step (\cref{fig:MS-SI-FIG-COHORT-FUNNEL}; \siref{MS-SI-WB-MATCHED-COHORT}). In order, the steps removed studies with no mutants, studies with no controls, Longshanks, treated animals (and any study then left without both groups), specimens not released on FaceBase, and specimens missing \texttt{Dataset}, \texttt{Strain}, \texttt{Sex}, or \texttt{Life\_Phase}, because a missing value that is more common in one group would itself predict the label. Each remaining specimen was assigned to a stratum, the combination of \texttt{Dataset}, \texttt{Strain}, and \texttt{Life\_Phase}, without merging distinct strain labels. Life phase is part of the stratum because several studies scanned controls and mutants at different ages, which would otherwise allow recorded variables to predict the label. Age enters only through life phase, so exact age remains a covariate within strata. Within each stratum and sex, the larger group was subsampled at random to the size of the smaller, and strata with fewer than five specimens per group were dropped. The result is 924 specimens in 16 strata from seven studies. A sensitivity variant stratifying on (\texttt{Dataset}, \texttt{Strain}) only is retained for comparison.

Two split designs were applied to the cohort, each grouped by biosample as for the existing splits. The first is five repeats of stratified 5-fold cross-validation (25 folds), in which every fold's test partition contains controls and mutants from every stratum in equal numbers, to within one specimen, with a stratified 20\% validation partition inside each training fold. The second is leave-one-study-out, with one partition for each of the seven studies. Automated checks assert that no biosample is in two partitions, that every stratum is in every cross-validation partition, and that controls and mutants stay balanced within each stratum. A balance check then tests whether the matched variables still predict the label. The metadata-only classifier is fit to \texttt{Dataset}, \texttt{Strain}, \texttt{Sex}, and \texttt{Life\_Phase}, and its out-of-fold balanced accuracy over all folds of each repeat is compared with a band of 0.45--0.55 around chance (0.5). Strata whose Clopper--Pearson interval excludes 0.5 are flagged.

The centroid-size and landmark classifiers were run on both split designs. The Stage 1 network was retrained on the same 25 folds and seven leave-one-study-out partitions at three training seeds, drawn from the operating-system random source and recorded before the corresponding runs began. The configuration of \emph{\secref{sec:network-training}} was unchanged, as an identical configuration hash across all 96 runs confirms. Class weights, early stopping, and checkpoint selection use only the split's training and validation partitions, and the test partition is scored once at the retained checkpoint. Per-specimen predictions, per-epoch curves, and a run log are retained for every run (\siref{MS-SI-WB-MATCHED-COHORT}). On the GPU used, cuDNN convolution kernels are nondeterministic and the training configuration does not disable them. Seed-to-seed spread therefore includes variation from nondeterministic GPU computation, and the checkpoint hashes identify the stored models rather than guaranteeing bit-for-bit reproduction.

Stratum-averaged aggregates are unweighted means over strata with a bootstrap over per-stratum values, and per-stratum AUCs are computed from the out-of-fold predictions combined across folds. The stratum-averaged AUCs quoted in the Results are means over the 25 folds of each fold's stratum average, whereas the per-stratum AUCs in \cref{fig:MS-FIG-MATCHED-CNN}A average each specimen's score over the five repeats before one AUC is computed per stratum. CNN metrics are computed in the same way as for the baselines, once on the 920 specimens the landmark classifiers scored (four cohort specimens have no landmark files), so that both are compared on identical specimens, and once on the full cohort of 924. Temperature scaling is fit per repeat on the validation predictions combined across folds and applied to the combined out-of-fold test predictions. A specimen tested in one fold is a validation specimen in others, so its label can enter the fit, and the calibrated values (\siref{MS-SI-WB-MATCHED-COHORT}) are not strictly out of sample. No value in the main text uses them.

\subsection{Statistical analysis}\label{sec:quantification-and-statistical-analysis}

\subsubsection{Sample sizes and significance}\label{sec:sample-sizes-and-significance}

Sample sizes are those the existing MusMorph partitions and the matched-cohort exclusion steps produced (\textit{n} = 422 and 132 test specimens; 924 cohort specimens). No power calculation was performed and no specimen was excluded from a CNN result other than by the split and cohort rules stated above. No hypothesis test is reported in the main text. Comparisons are reported as point estimates with 95\% intervals, and the count of errors within study and strain is compared with its expectation under random assignment (\emph{\nameref{sec:statistical-analysis-of-error-structure}}). Exact \textit{n} for every estimate is given in the corresponding table.

\subsubsection{Confidence intervals}\label{sec:confidence-intervals}

Confidence intervals on primary metrics are from 10,000 stratified bootstrap resamples of the test-set predictions (each class resampled to its observed count, the same indices used across metrics within a stage), reported as the 2.5th and 97.5th percentiles. Because stratified resampling yields zero-width intervals for classes with no observed errors, per-class recall, within-stratum accuracies, and subset specificities use Clopper--Pearson exact intervals.

\subsubsection{Calibration}\label{sec:calibration}

Reliability diagrams and the expected calibration error (ECE) use bins of the predicted-class probability. Stage 1 uses five equal-count quantile bins, with tied edges merged, because Stage 1 confidences cluster near 1 and equal-width bins would place most predictions in the top bin. Stage 2 uses ten bins of equal width, the conventional scheme,\cite{guo2017} with empty bins skipped. The ECE is the count-weighted mean absolute gap between mean confidence and empirical accuracy across bins. Temperature scaling\cite{guo2017} was fit separately for each stage by minimizing validation negative log-likelihood over a scalar $T \in [0.1, 50]$ applied to the logits, which were recovered as the log of the saved softmax probabilities (exact up to an additive constant the softmax removes; no saved probability is zero). The fitted $T$ was applied to the test logits and ECE recomputed with the same binning rule. Temperature scaling is monotone in the logits and cannot change the argmax, and we verified that no test prediction changed.

\subsubsection{Error structure}\label{sec:statistical-analysis-of-error-structure}

Each of the 20 test errors was classified by whether its predicted label is recorded in the specimen's study and in its stratum (the study, with the Enhancer study divided by background strain). The number of such errors expected if each error fell at random on one of the other 15 labels is the sum, over errors, of the number of other labels in the true label's stratum (or study) divided by 15, which gives 2.2 for the stratum and 5.9 for the study. Of the 120 unordered pairs of labels, eight share a stratum (\siref{MS-SI-TAB-ERROR-PROVENANCE}; \siref{MS-SI-TAB-LABEL-PAIRS}). Statements about biological relationships among labels are taken from the publications of the contributing studies and were not assessed independently.

Per-class asymmetries were checked for consistency across partitions by comparing per-class recall on the test and validation predictions of the same model (Diagnostic 6). Because both partitions come from the same sources, this checks consistency, not independent replication.

\subsubsection{Paired comparison on the matched cohort}\label{sec:paired-comparison-on-the-matched}

The CNN$-$baseline difference on the matched cohort was computed from the stored out-of-fold predictions of both classifiers on the same 25 folds and the 920 commonly scored specimens, for each CNN seed and each baseline classifier. For each stratum, AUC was averaged over the 25 folds and the CNN$-$baseline difference taken. For the cranium-and-mandible landmark classifier, the difference was also computed in balanced accuracy. These per-stratum differences were averaged over the 16 strata, with a percentile interval from 10,000 bootstrap resamples of the strata and the count of positive strata. Because strata are nested within studies and resampled independently, the stratum bootstrap describes variation across the represented strata rather than across studies. Two secondary forms, a stratum bootstrap on each specimen's score averaged over the five repeats and a repeat-level $t$-interval on four degrees of freedom, are in \siref{MS-SI-WB-MATCHED-COHORT}.

\subsection{Computational reproducibility}\label{sec:computational-reproducibility}

We developed the project with the CASTER template,\cite{schuler2026caster} a reusable project scaffold, distinct from this project's code, that applies the continuous-FAIRness approach of Deriva-ML\cite{li2024deriva} to machine-learning projects run by small teams. Records of how each result was produced are therefore captured while the work is done rather than reconstructed afterward. All code and configuration are version-controlled, and each configurable parameter is defined in one configuration file. The source data files used were recorded, and the derived datasets, data splits, and trained models are tracked by content hash (a fingerprint of each file's contents). Every run records its code version, configuration, input data, software environment, completion status, and primary metrics under version control, so each reported result can be traced to the run that produced it. Every random seed was drawn from the operating-system random source and recorded. Every figure and table was regenerated from stored run outputs and linked to the runs that produced it. Anthropic's Claude models were used as AI coding assistants, and the authors reviewed all code.

Data preparation ran on an Amazon Web Services (AWS) m7i.2xlarge instance, and CNN training and evaluation on an AWS g6.xlarge instance with one NVIDIA L4 GPU. Cohort construction, the centroid-size and landmark classifiers, and figure generation ran on an Apple M2 workstation. A version-controlled lockfile fixes the version of every Python dependency, including PyTorch. Because dependencies were updated as the project evolved, the exact package versions behind each result are those recorded for the run that produced it.

\subsection{Key resources table}\label{sec:key-resources-table}

{\small\begin{longtable}{@{}p{5.5cm}p{4.4cm}p{5.5cm}@{}}
\toprule
\textbf{Reagent or resource} & \textbf{Source} & \textbf{Identifier} \\
\midrule
\endfirsthead
\toprule
\textbf{Reagent or resource} & \textbf{Source} & \textbf{Identifier} \\
\midrule
\endhead
\multicolumn{3}{@{}l}{\textbf{Deposited data}} \\
MusMorph adult mouse skull micro-CT volumes, sparse and dense landmarks, and specimen metadata (MusMorph project) & Devine et al.;\cite{devine2022} Hallgr{\'i}msson et al.\cite{musmorph_project} & FACEBASE:3-HXMC \\
MusMorph project-wide specimen and study metadata (\texttt{MusMorph\_Metadata.csv}, \texttt{Study\_Metadata.csv}) & Hallgr{\'i}msson et al.\cite{musmorph_projectwide} & FACEBASE:6-2EPY \\
MusMorph B9d dataset & Hallgr{\'i}msson et al.\cite{musmorph_b9d} & FACEBASE:3-JQMM \\
MusMorph Bmp dataset & Hallgr{\'i}msson et al.\cite{musmorph_bmp} & FACEBASE:3-KB46 \\
MusMorph Brachymorph dataset & Hallgr{\'i}msson et al.\cite{musmorph_brachymorph} & FACEBASE:3-KB1W \\
MusMorph Brain-Face (\texttt{Brain\_Face}) dataset & Hallgr{\'i}msson et al.\cite{musmorph_brainface} & FACEBASE:3-KB3J \\
MusMorph Bulgy dataset & Hallgr{\'i}msson et al.\cite{musmorph_bulgy} & FACEBASE:3-JZ9G \\
MusMorph Collaborative Cross (\texttt{CC}) dataset & Hallgr{\'i}msson et al.\cite{musmorph_collaborativecross} & FACEBASE:3-KB0W \\
MusMorph Diversity Outbred (\texttt{DO}) dataset & Hallgr{\'i}msson et al.\cite{musmorph_diversityoutbred} & FACEBASE:3-KB4P \\
MusMorph Enhancer dataset & Hallgr{\'i}msson et al.\cite{musmorph_enhancer} & FACEBASE:3-KB02 \\
MusMorph Fgf dataset & Hallgr{\'i}msson et al.\cite{musmorph_fgf} & FACEBASE:3-JVEM \\
MusMorph Ghrhr dataset & Hallgr{\'i}msson et al.\cite{musmorph_ghrhr} & FACEBASE:3-KB08 \\
MusMorph Hybrid dataset & Hallgr{\'i}msson et al.\cite{musmorph_hybrid} & FACEBASE:3-KB32 \\
MusMorph Ift dataset & Hallgr{\'i}msson et al.\cite{musmorph_ift} & FACEBASE:3-JVE0 \\
MusMorph Longshanks dataset & Hallgr{\'i}msson et al.\cite{musmorph_longshanks} & FACEBASE:3-KFBE \\
MusMorph Lrp dataset & Hallgr{\'i}msson et al.\cite{musmorph_lrp} & FACEBASE:3-KB4J \\
MusMorph MPS dataset & Hallgr{\'i}msson et al.\cite{musmorph_mps} & FACEBASE:3-KFBY \\
MusMorph Nipbl dataset & Hallgr{\'i}msson et al.\cite{musmorph_nipbl} & FACEBASE:3-KB0J \\
MusMorph Nosip dataset & Hallgr{\'i}msson et al.\cite{musmorph_nosip} & FACEBASE:3-JQMP \\
MusMorph Osteo Imperfecta (\texttt{Osteo\_Imperfecta}) dataset & Hallgr{\'i}msson et al.\cite{musmorph_osteoimperfecta} & FACEBASE:3-KB00 \\
MusMorph Pten dataset & Hallgr{\'i}msson et al.\cite{musmorph_pten} & FACEBASE:3-KFZJ \\
MusMorph Shh dataset & Hallgr{\'i}msson et al.\cite{musmorph_shh} & FACEBASE:3-JVE6 \\
MusMorph Spry dataset & Hallgr{\'i}msson et al.\cite{musmorph_spry} & FACEBASE:3-JZAM \\
MusMorph Strain Comparison (\texttt{Strain\_Comparison}) dataset & Hallgr{\'i}msson et al.\cite{musmorph_straincomparison} & FACEBASE:3-JZ9J \\
MusMorph Trp dataset & Hallgr{\'i}msson et al.\cite{musmorph_trp} & FACEBASE:3-KFZW \\
Derived data: split tables, matched-cohort definition and splits, per-specimen predictions for every model and baseline, bootstrap draws, preprocessed 48$^{3}$ volumes, and the trained Stage 1, Stage 2, and matched-cohort model checkpoints & This paper & Public repository; identifier on journal publication \\
Run records and the source tables for every figure and supplemental table & This paper & \url{https://github.com/robes/mapl} \ifpreprint(public on journal publication)\else(private during review; public at publication)\fi \\
\midrule
\multicolumn{3}{@{}l}{\textbf{Software and algorithms}} \\
Analysis code for both classifiers, provenance analysis, calibration, baselines, matched cohort and figures & This paper & \url{https://github.com/robes/mapl} \ifpreprint(public on journal publication)\else(private during review; public at publication)\fi \\
MusMorph preprocessing and landmarking pipeline & Devine et al.\cite{devine2022} & \url{https://github.com/jaydevine/MusMorph} \\
Python 3.12 (versions of all Python packages below are pinned in the repository lockfile and recorded per run in its provenance record) & Python Software Foundation & RRID:SCR\_008394; \url{https://www.python.org} \\
PyTorch (lockfile-pinned) & PyTorch Foundation & RRID:SCR\_018536; \url{https://pytorch.org} \\
NumPy (lockfile-pinned) & NumPy developers & RRID:SCR\_008633; \url{https://numpy.org} \\
SciPy (lockfile-pinned) & SciPy developers & RRID:SCR\_008058; \url{https://scipy.org} \\
scikit-learn (lockfile-pinned) & scikit-learn developers & RRID:SCR\_002577; \url{https://scikit-learn.org} \\
NiBabel (lockfile-pinned) & NIPY & \url{https://nipy.org/nibabel} \\
DVC (data version control) (lockfile-pinned) & Iterative & \url{https://dvc.org} \\
CASTER project template (continuous-FAIRness provenance capture) & Schuler\cite{schuler2026caster} & \url{https://github.com/robes/caster} \\
Project development container (Docker; Debian 12, Python 3.12) & This paper & Dockerfile in \url{https://github.com/robes/mapl} \\
\midrule
\multicolumn{3}{@{}l}{\textbf{Other}} \\
VISTA Enhancer Browser (naming convention for \emph{hs1629}) & Visel et al.\cite{visel2007} & \url{https://enhancer.lbl.gov} \\
\bottomrule
\end{longtable}}

\section{Resource availability}\label{sec:resource-availability}

\subsection{Lead contact}\label{sec:lead-contact}

Further information and requests for resources should be directed to and will be fulfilled by the lead contact, Robert Schuler (schuler@isi.edu).

\subsection{Materials availability}\label{sec:materials-availability}

This study did not generate new unique reagents. No new animal work was performed.

\subsection{Data and code availability}\label{sec:data-and-code-availability}

This paper analyzes existing, publicly available data\cite{musmorph_b9d,musmorph_bmp,musmorph_brachymorph,musmorph_brainface,musmorph_bulgy,musmorph_collaborativecross,musmorph_diversityoutbred,musmorph_enhancer,musmorph_fgf,musmorph_ghrhr,musmorph_hybrid,musmorph_ift,musmorph_longshanks,musmorph_lrp,musmorph_mps,musmorph_nipbl,musmorph_nosip,musmorph_osteoimperfecta,musmorph_pten,musmorph_shh,musmorph_spry,musmorph_straincomparison,musmorph_trp,musmorph_projectwide} from MusMorph on FaceBase under project FACEBASE:3-HXMC.\cite{musmorph_project} The identifiers of the datasets used are listed in the key resources table. The derived data, including the data splits, per-specimen predictions, preprocessed volumes, and trained models, will be deposited in a public repository and released on the date of publication.

All original code, together with the configuration files, run records, and source tables for every figure and supplemental table, is in the repository \url{https://github.com/robes/mapl}. \ifpreprint\ifpreprintpubliccode The repository is public, and the version described here is tagged \texttt{\preprinttag}. A further tagged commit will be listed on the date of publication.\else The repository will be released publicly on journal publication.\fi\else The repository is private during peer review, is available to reviewers on request through the editor, and will be released publicly on the date of publication.\fi{}

Any additional information required to reanalyze the data reported in this paper is available from the lead contact upon request.

%% file: sections/06_backmatter.tex
\section{Acknowledgments}\label{sec:acknowledgments}

This work was supported in part by the National Institute of Dental and Craniofacial Research, the National Institute on Deafness and Other Communication Disorders, and the NIH Office of Data Science Strategy under grant U24DE034163. The funders had no role in study design, data collection and analysis, decision to publish, or preparation of the manuscript. The data analyzed here were obtained from the FaceBase repository (\url{https://www.facebase.org}). We thank the MusMorph authors and the contributing laboratories for making the specimens and their study-level metadata publicly available. We thank Jifan Feng for reviewing the manuscript and for constructive comments that improved it.

\section{Author contributions}\label{sec:author-contributions}

Conceptualization, R.S.; data curation, R.S.; formal analysis, R.S. and C.S.; investigation, R.S. and C.S.; methodology, R.S. and C.S.; software, R.S. and C.S.; validation, R.S. and C.S.; visualization, R.S.; writing -- original draft, R.S. and C.S.; writing -- review \& editing, R.S. and C.S.

\section{Declaration of interests}\label{sec:declaration-of-interests}

R.S. is a member of the FaceBase team and a co-author of the FaceBase resource papers cited here.\cite{schuler2022facebase,samuels2020facebase} R.S. is also a co-author of the MusMorph data descriptor\cite{devine2022} and a contributor to the MusMorph project and datasets on FaceBase.\cite{musmorph_project} The authors declare no other competing interests.

\section{Declaration of generative AI and AI-assisted technologies in the manuscript preparation process}\label{sec:declaration-of-generative-ai-and}

During the preparation of this work, the authors used Claude (Anthropic) to help draft and edit text, and ChatGPT (OpenAI) and Gemini (Google) to review drafts for readability and suggest revisions. After using these tools, the authors reviewed and edited the content as needed and take full responsibility for the content of the published article.

\section{Supplemental information}\label{sec:supplemental-information}

\ifpreprint\siref{MS-SI-DOCUMENT}, with \crefrange{fig:MS-FIG-ARCHITECTURE}{fig:MS-SI-FIG-COHORT-FUNNEL}, follows the references. \sirange{MS-SI-WB-STAGE2-PREDICTION-PROVENANCE}{MS-SI-WB-MATCHED-COHORT} accompany this preprint as Excel workbooks, one sheet per source table.\else Supplemental information can be found online at [DOI to be inserted by the publisher]. \siref{MS-SI-DOCUMENT} contains \crefrange{fig:MS-FIG-ARCHITECTURE}{fig:MS-SI-FIG-COHORT-FUNNEL}. \sirange{MS-SI-WB-STAGE2-PREDICTION-PROVENANCE}{MS-SI-WB-MATCHED-COHORT} are provided as Excel workbooks, one sheet per source table.\fi{} Each item's title and legend, with the main-text figure or table it relates to, is given in \siref{MS-SI-DOCUMENT}.

%% file: sections/08_document_s1.tex
\clearpage
\section*{\siname{MS-SI-DOCUMENT}. Supplemental information}\phantomsection\label{text:MS-SI-DOCUMENT}
\setcounter{figure}{0}
\renewcommand{\thefigure}{S\arabic{figure}}

\noindent\textit{\sirange{MS-SI-WB-STAGE2-PREDICTION-PROVENANCE}{MS-SI-WB-MATCHED-COHORT} are Excel workbooks, one sheet per source table. Each legend below names the sheets. Most workbook columns use MusMorph field names. \texttt{Dataset} is the contributing study, \texttt{Experimental\_\allowbreak Group} is the control or mutant designation, and \texttt{Gene} is the gene label. Some sheets name the study and gene label \texttt{study} and \texttt{label}, or give a field name in lowercase. Columns named \texttt{strain\_group} identify a label's Stage 2 stratum: its background strain within the Enhancer study, or all strains within Spry (\methodsref{sec:study-and-strain-decomposition}).}

\begin{figure}[H]
\centering
\includegraphics[width=0.92\textwidth,height=0.8\textheight,keepaspectratio]{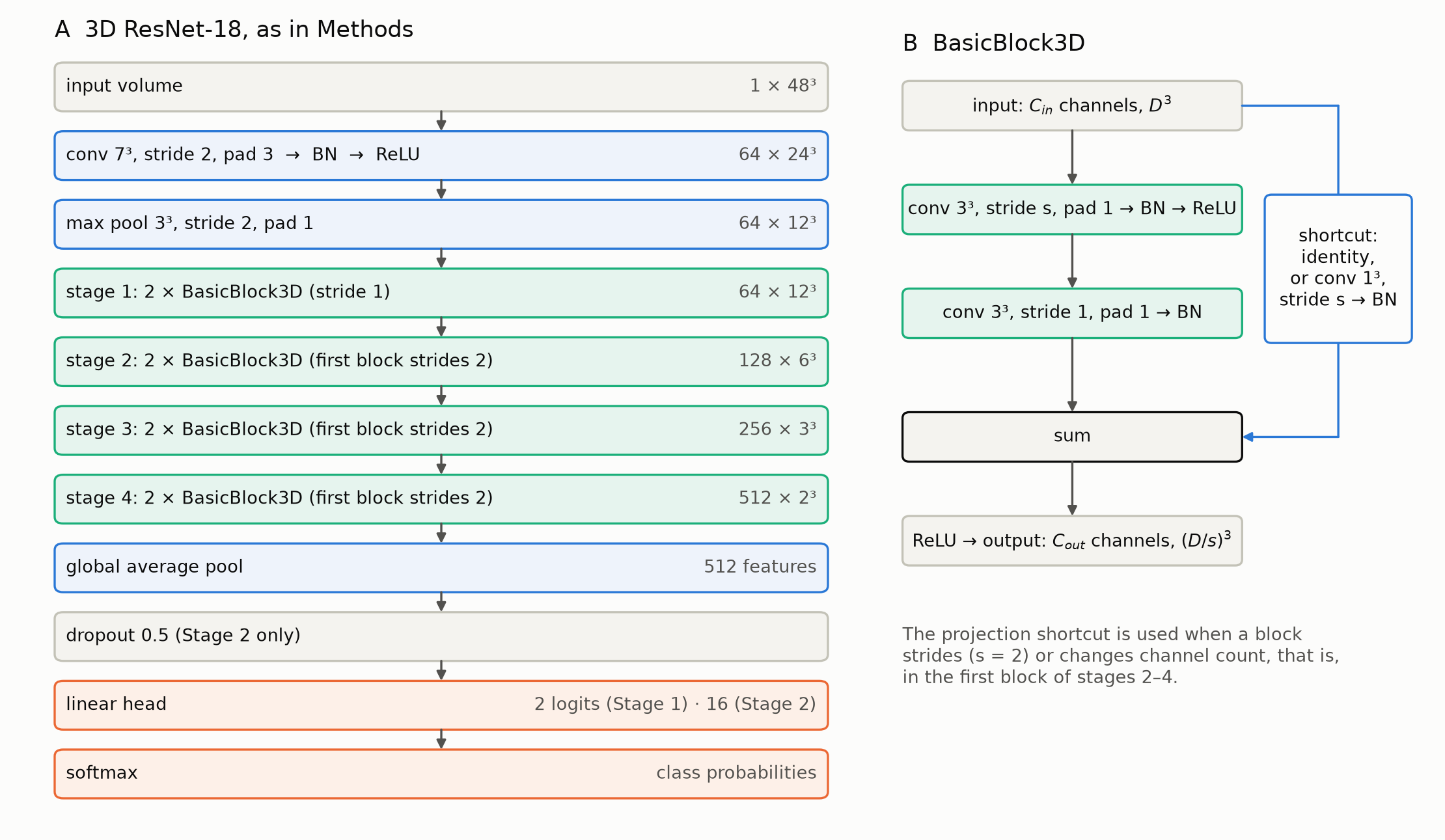}
\caption{\textbf{3D ResNet-18 backbone shared by both classifiers. Related to \cref{fig:MS-FIG-CONVENTIONAL} and \methodsref{sec:network-training}.} A single-channel 48$^{3}$ volume passes through a 7 $\times$ 7 $\times$ 7 stride-2 convolution, BatchNorm, ReLU, and 3 $\times$ 3 $\times$ 3 stride-2 max-pooling (64 feature maps at 12$^{3}$), then four residual stages of two BasicBlock3D units each (64 $\rightarrow$ 128 $\rightarrow$ 256 $\rightarrow$ 512 channels at 12$^{3}$, 6$^{3}$, 3$^{3}$, and 2$^{3}$), global average pooling, and a linear head (two logits at Stage 1; dropout p = 0.5 and 16 logits at Stage 2); inset, the BasicBlock3D unit with identity or 1 $\times$ 1 $\times$ 1 projection shortcut. Spatial sizes were verified by forward hooks on the implemented model with the trained weights loaded.}
\label{fig:MS-FIG-ARCHITECTURE}
\end{figure}

\begin{figure}[H]
\centering
\includegraphics[width=0.92\textwidth,height=0.8\textheight,keepaspectratio]{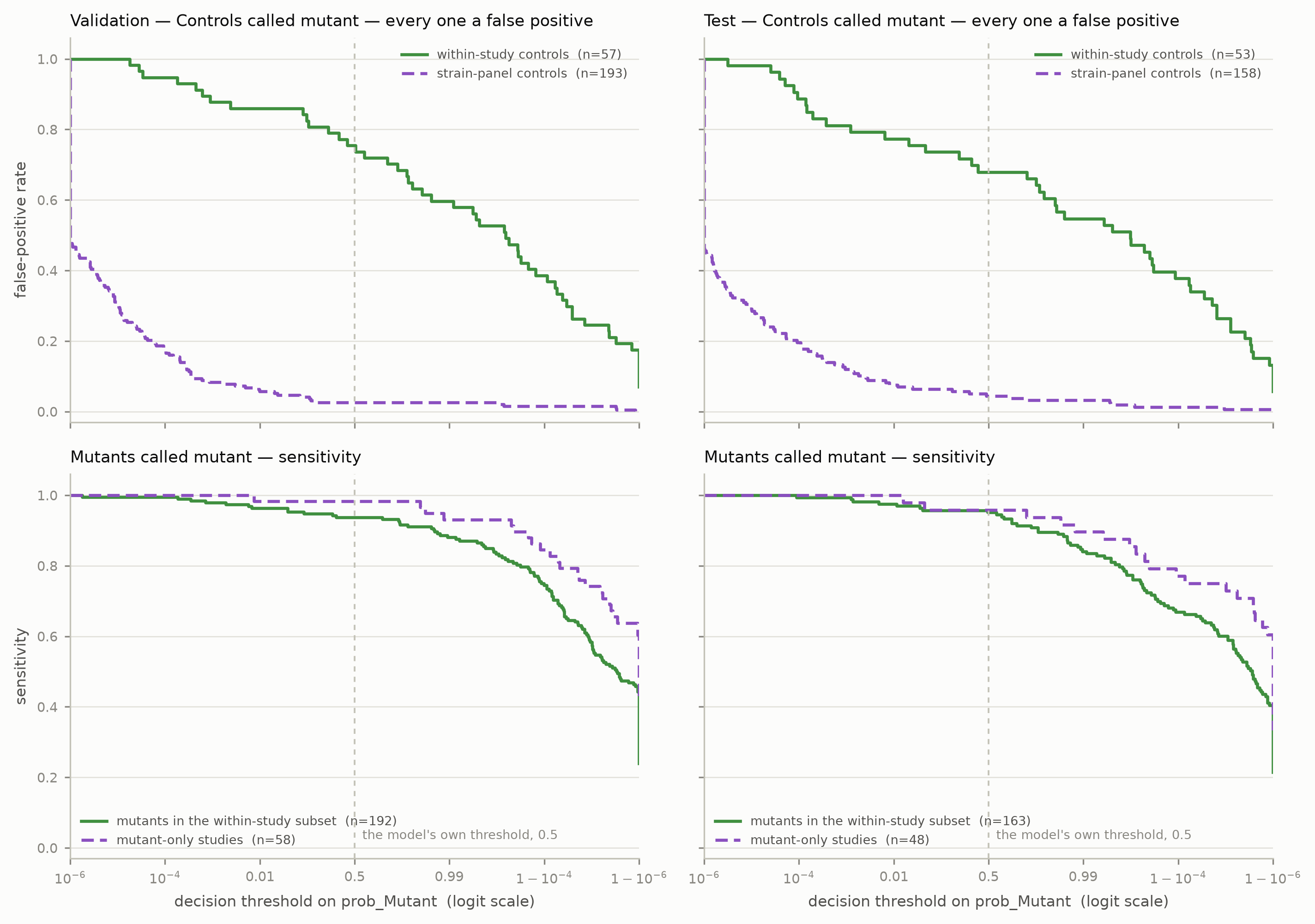}
\caption{\textbf{Stage 1 decision threshold swept by provenance. Related to \cref{fig:MS-FIG-CONVENTIONAL,fig:MS-FIG-STAGE1-PROVENANCE}.} Stage 1 decision threshold swept on a logit axis from 10$^{-6}$ to 1 $-$ 10$^{-6}$, on the validation (left) and test (right) partitions, with the false-positive rate among controls split by provenance (within-study controls vs.\ strain-panel controls; top) and sensitivity among mutants split the same way (bottom). In the within-study subset, the false-positive rate among controls and the sensitivity among mutants decline together as the threshold rises. The strain-panel false-positive rate is near zero throughout. Operating points chosen on validation for target specificities of 0.80, 0.90, and 0.95 on the within-study subset require thresholds of 0.9999983, exactly 1.0, and above 1.0 (that is, unattainable), respectively.}
\label{fig:MS-SI-FIG-THRESHOLD-SWEEP}
\end{figure}

\begin{figure}[H]
\centering
\includegraphics[width=0.92\textwidth,height=0.8\textheight,keepaspectratio]{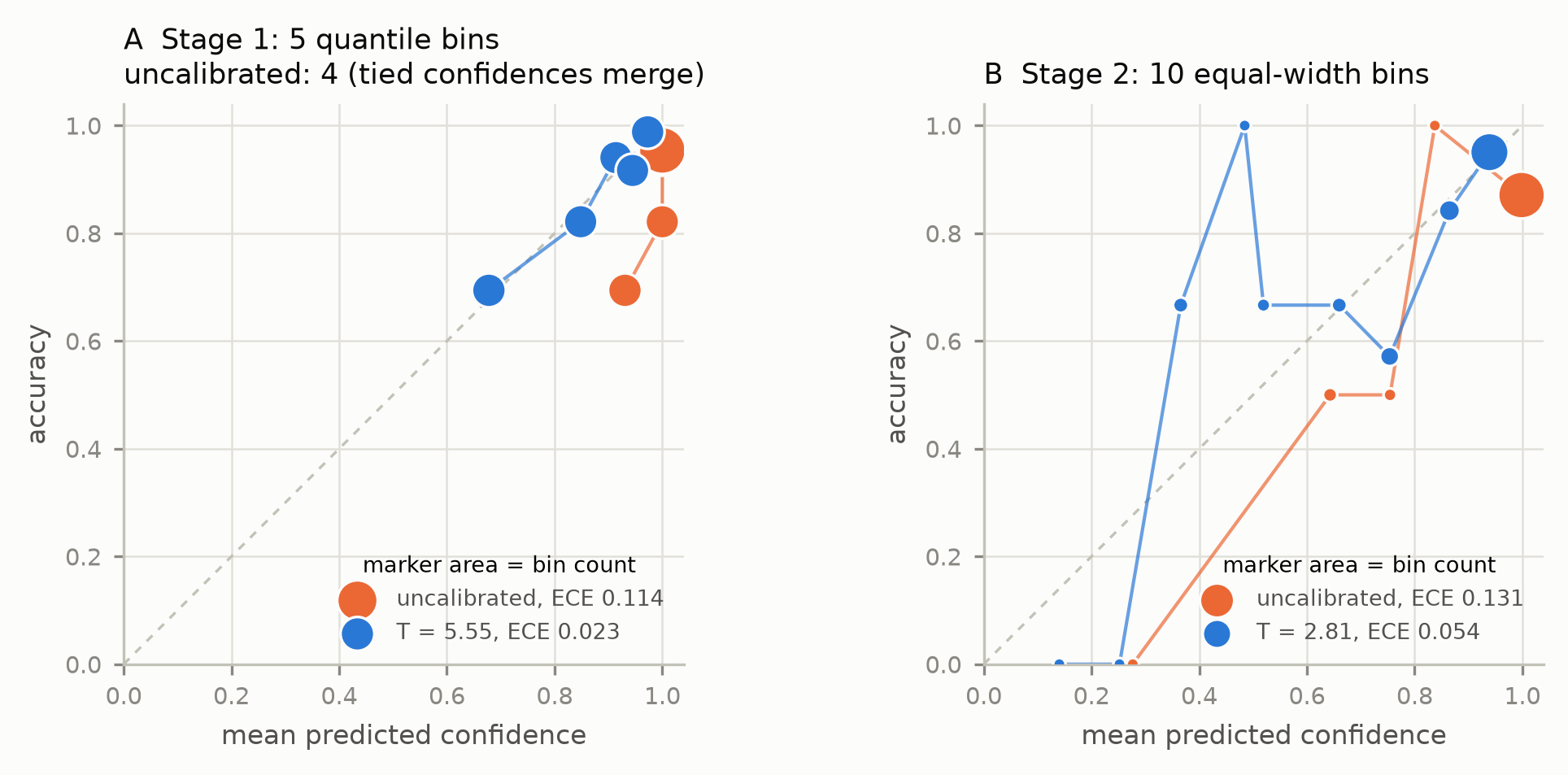}
\caption{\textbf{Reliability diagrams for both stages before and after temperature scaling. Related to \cref{fig:MS-FIG-CONVENTIONAL} and \siref{MS-SI-WB-INTERVALS-CALIBRATION}.} Marker area is proportional to bin count. Stage 1 uses five equal-count bins, which reduce to four before scaling because the two highest bin edges tie at a predicted probability of 1.0. Stage 2 uses ten equal-width bins, with empty bins skipped. Every uncalibrated bin lies below the diagonal at Stage 1 (overconfident). Scaling brings both stages close to the diagonal without changing any prediction.}
\label{fig:MS-SI-FIG-RELIABILITY}
\end{figure}

\begin{figure}[H]
\centering
\includegraphics[width=0.92\textwidth,height=0.8\textheight,keepaspectratio]{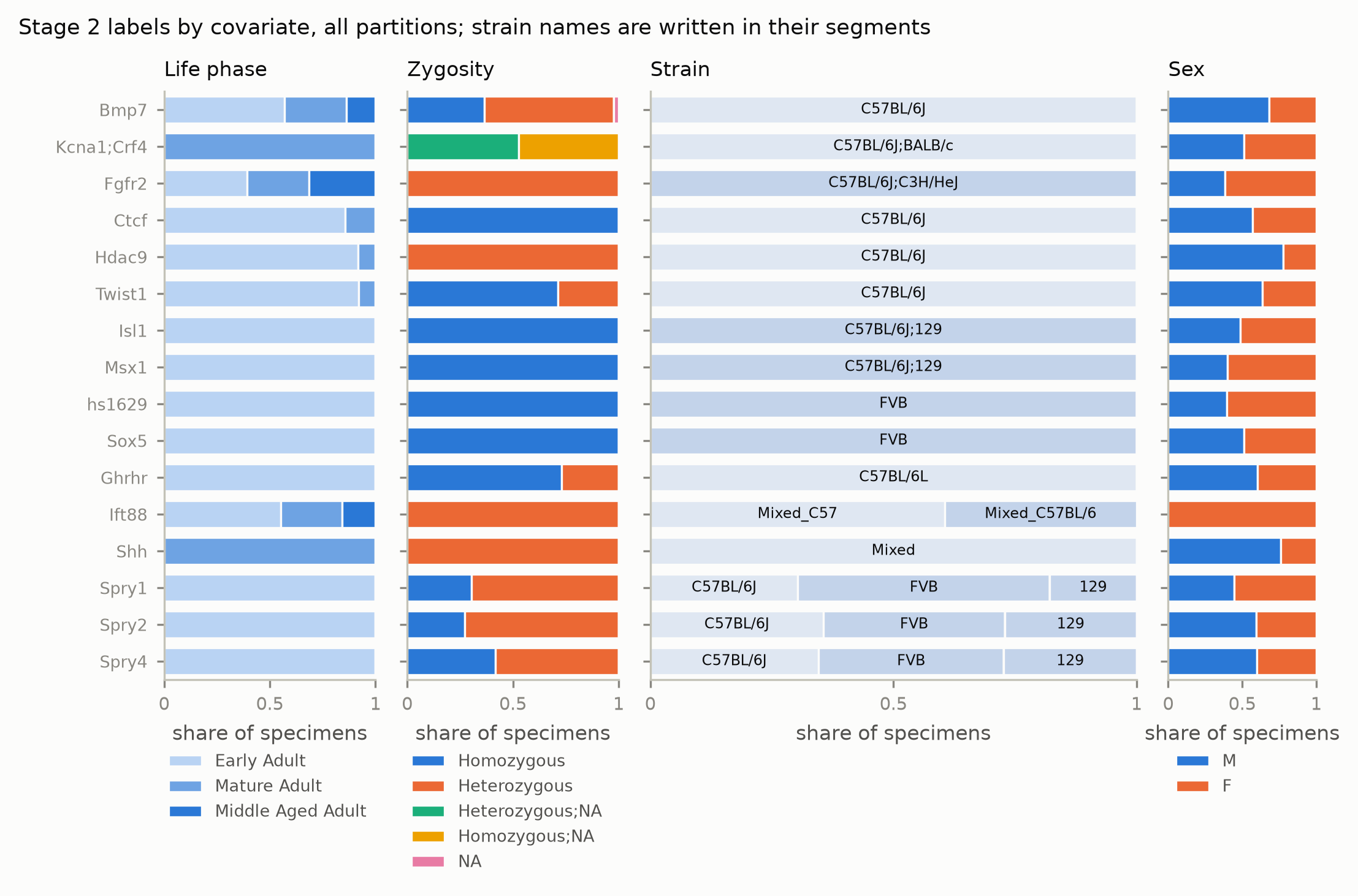}
\caption{\textbf{Composition of the 16 gene labels by covariate. Related to \cref{fig:MS-FIG-PROVENANCE-STRUCTURE}.} Labels are broken down by life phase, zygosity, strain, and sex across all partitions (strain names written in their segments as recorded). Life phase, strain, and zygosity are largely determined by the label.}
\label{fig:MS-SI-FIG-STAGE2-COMPOSITION}
\end{figure}

\begin{figure}[H]
\centering
\includegraphics[width=0.92\textwidth,height=0.8\textheight,keepaspectratio]{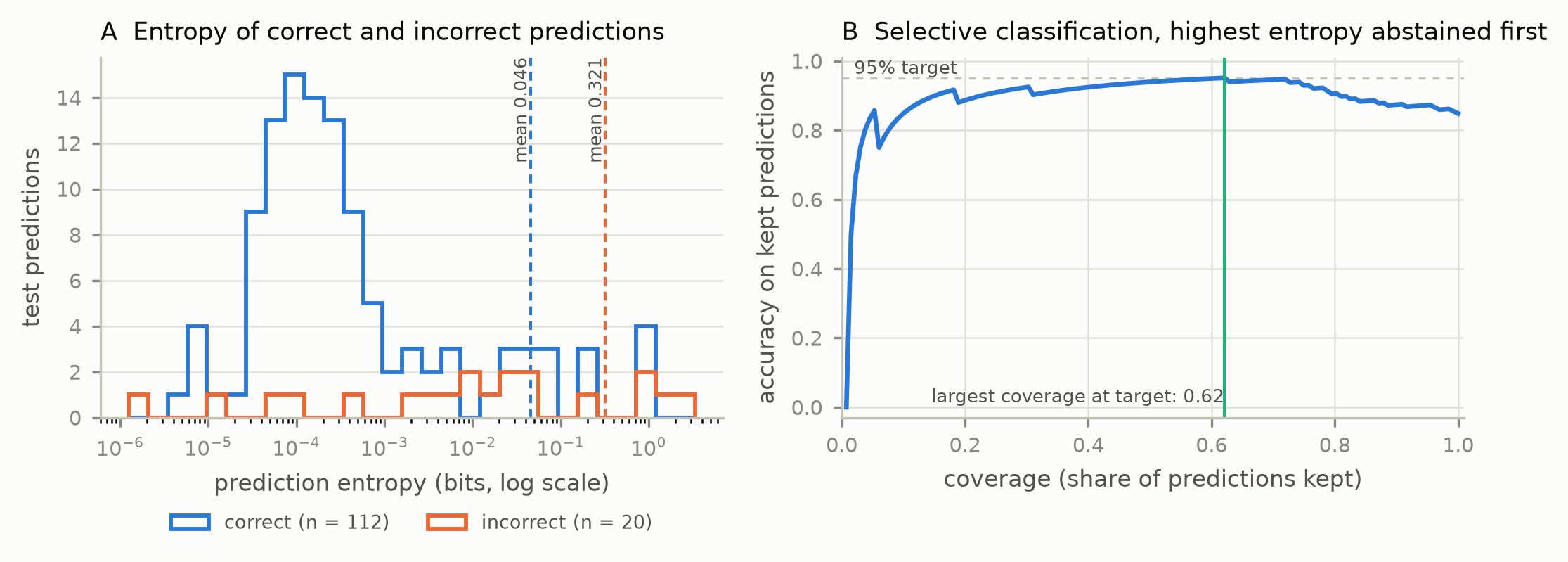}
\caption{\textbf{Stage 2 uncertainty and selective classification. Related to \cref{fig:MS-FIG-STAGE2-ERRORS}.} All panels use the 132-specimen test set and the uncalibrated softmax. (A) Distribution of Shannon prediction entropy for correct (\textit{n} = 112) and incorrect (\textit{n} = 20) predictions. The heavy right tail of the error distribution consists of the two cross-study errors (1.43 and 2.64 bits), and the 18 same-study errors have a median entropy of 0.010 bits. (B) Accuracy on the retained subset as a function of coverage when the highest-entropy predictions are withheld first. Retained accuracy reaches 0.95 only at 62\% coverage and is not monotone, because same-study errors are made at very low entropy and remain after withholding.}
\label{fig:MS-FIG-STAGE2-UNCERTAINTY}
\end{figure}

\begin{figure}[H]
\centering
\includegraphics[width=0.92\textwidth,height=0.8\textheight,keepaspectratio]{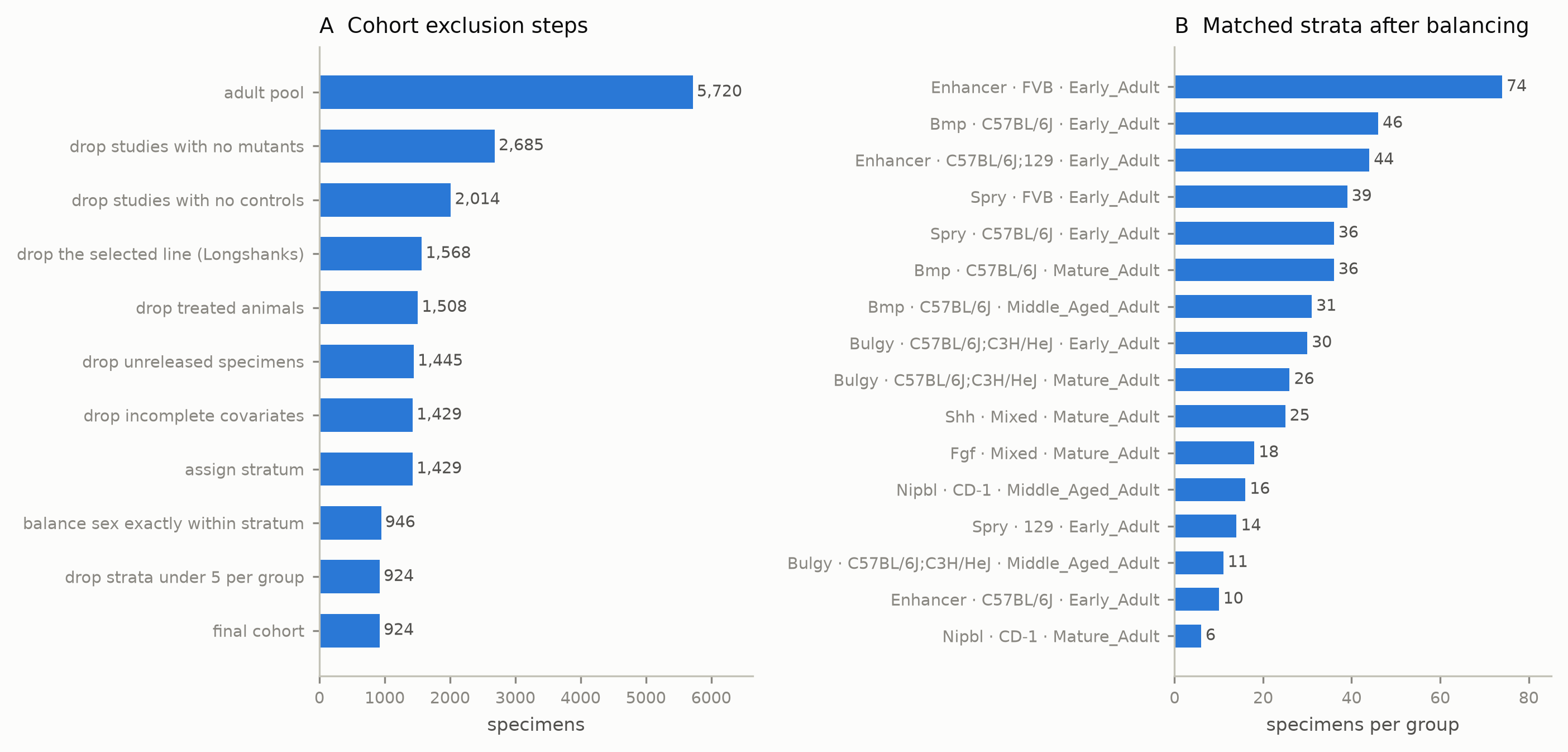}
\caption{\textbf{Matched-cohort flow diagram and per-stratum counts. Related to \cref{fig:MS-FIG-MATCHED-CNN} and \siref{MS-SI-WB-MATCHED-COHORT}.} (A) Specimens remaining after each exclusion step, from the adult skull pool restricted to controls and mutants (5,720) to the final cohort of 924. (B) The 16 strata after balancing. Each stratum is a study $\times$ strain $\times$ life phase combination, and study names (for example Bmp) are MusMorph \texttt{Dataset} values, not gene labels. Bars give the number of specimens per group. Every stratum holds equal numbers of controls and mutants, and the task within it is control-versus-mutant classification.}
\label{fig:MS-SI-FIG-COHORT-FUNNEL}
\end{figure}

\subsection*{Supplemental tables (Excel workbooks)}
\noindent Results for the diagnostics in \cref{tab:MS-TAB-DIAGNOSTICS} are in the workbooks below.

\begin{center}
\small
\begin{tabular}{@{}ll@{}}
\toprule
\textbf{Diagnostic} & \textbf{Workbook (sheets)} \\
\midrule
1 Label--source nesting & \siref{MS-SI-WB-STAGE2-SPLIT-PROVENANCE}; \siref{MS-SI-WB-STAGE1-SPLIT-PROVENANCE} (A, C); \siref{MS-SI-WB-INTERVALS-CALIBRATION} (D) \\
2 Provenance-only baseline & \siref{MS-SI-WB-INTERVALS-CALIBRATION} (A, D); \siref{MS-SI-WB-BASELINES} (A, C) \\
3 Within-source evaluation & \siref{MS-SI-WB-STAGE2-PREDICTION-PROVENANCE} (I); \siref{MS-SI-WB-STAGE1-PREDICTION-PROVENANCE} (C, D); \siref{MS-SI-WB-BASELINES} (A, C) \\
4 Own-source concentration & \siref{MS-SI-WB-STAGE2-PREDICTION-PROVENANCE} (B, I) \\
5 Error-pattern check & \siref{MS-SI-WB-STAGE2-PREDICTION-PROVENANCE} (G, H) \\
6 Cross-partition consistency & \siref{MS-SI-WB-STAGE2-PREDICTION-PROVENANCE} (F); \siref{MS-SI-WB-INTERVALS-CALIBRATION} (A) \\
\bottomrule
\end{tabular}
\end{center}

\begin{itemize}

\item\phantomsection\label{tab:MS-SI-WB-STAGE2-PREDICTION-PROVENANCE}\textbf{\siname{MS-SI-WB-STAGE2-PREDICTION-PROVENANCE}. Stage 2 prediction-level provenance. Related to \cref{fig:MS-FIG-STAGE2-ERRORS}.} Sheets: (A) The 20 Stage 2 errors with study, study-and-strain stratum, zygosity, and entropy. (B) Study-vs-gene decomposition and per-study accuracy with own-study probability. (C) Spry correctness by gene, strain, and zygosity; Enhancer correctness by background strain and gene. (D) Prediction entropy by outcome (correct, same-study error, cross-study error). (E) Joint performance of the two stages on the 132 shared test specimens, counting a specimen as correct only if Stage 1 calls it mutant and Stage 2 assigns its gene label correctly, at three Stage 1 thresholds, and the Stage 1 misses among these specimens. (F) Stage 2 per-class precision, recall (with Clopper--Pearson 95\% CI), F1, and number of test specimens on the 132-specimen test set, with the contributing study and background strain of each class. (G) The 20 errors, each marked by whether its predicted label is recorded in the specimen's study and in its study-and-strain stratum, with the number of other labels there; and the observed counts of such errors beside the counts expected if each error fell at random on one of the other 15 labels. (H) The 120 unordered pairs of labels, each marked by whether the two labels share a study and a stratum, with the number of errors on the pair; and a summary by whether the pair shares a stratum. (I) Decomposition of Stage 2 accuracy into study-level and within-study components, with Clopper--Pearson CIs.

\item\phantomsection\label{tab:MS-SI-WB-INTERVALS-CALIBRATION}\textbf{\siname{MS-SI-WB-INTERVALS-CALIBRATION}. Primary-metric intervals, provenance-only baselines, and calibration. Related to \cref{fig:MS-FIG-CONVENTIONAL}.} Sheets: (A) Bootstrap 95\% confidence intervals on all primary metrics; validation-set per-class recall for Stage 2; majority-class baselines by provenance group (labeled ``training majority''); temperature-scaling fits; and additional intervals for subset metrics reported in the main text, with method and counts. (B) Post hoc temperature scaling for both stages: fitted temperature, test negative log-likelihood and ECE before and after scaling, and the number of predictions changed (zero at both stages). (C) Conventional test performance of the two stages with 95\% bootstrap CIs and ECE. (D) Study and strain structure of the labels, baselines given each specimen's true stratum (labeled ``oracle''), and fitted metadata-only classifiers. In sheet D, ``control-only panels'' are the strain panels.

\item\phantomsection\label{tab:MS-SI-WB-STAGE2-SPLIT-PROVENANCE}\textbf{\siname{MS-SI-WB-STAGE2-SPLIT-PROVENANCE}. Stage 2 split-level provenance. Related to \cref{fig:MS-FIG-PROVENANCE-STRUCTURE}.} Sheets: (A) Gene $\times$ contributing study, all partitions and test only. (B) Per-gene provenance: study, strains, genotypes, zygosity, treatments, life phases, partition counts, co-resident genes, sole-gene flag. (C) Gene $\times$ partition by strain, zygosity, life phase, and sex. (D) Information summary: H(Gene), H(Gene $\mid$ Dataset), accuracy of the baselines given the true stratum. (E) Enhancer study: strain $\times$ gene $\times$ genotype.

\item\phantomsection\label{tab:MS-SI-WB-STAGE1-SPLIT-PROVENANCE}\textbf{\siname{MS-SI-WB-STAGE1-SPLIT-PROVENANCE}. Stage 1 split-level provenance. Related to \cref{fig:MS-FIG-PROVENANCE-STRUCTURE}.} Sheets: (A) Stage 1 study $\times$ partition $\times$ group; training control fraction per study; test-control provenance. (B) Full MusMorph adult skull pool by study. (C) Strain $\times$ (group, study); mutant gene $\times$ zygosity; Stage 1 information summary. Sheets A (study by partition and group) and C (strain by group and study) have two header rows, and the first column holds the row labels. Control-only studies (flagged \texttt{control\_only\_dataset}) are the strain panels (CC, DO, Hybrid, Strain\_Comparison).

\item\phantomsection\label{tab:MS-SI-WB-STAGE1-PREDICTION-PROVENANCE}\textbf{\siname{MS-SI-WB-STAGE1-PREDICTION-PROVENANCE}. Stage 1 prediction-level provenance. Related to \cref{fig:MS-FIG-STAGE1-PROVENANCE}.} Sheets: (A) False positives by study (with count at mutant probability $>$ 0.99) and within-study controls by strain. (B) False negatives with gene, zygosity, genotype, and strain; mutant recall by study. (C) Stage 1 metrics by provenance subset and per-study AUC in the within-study subset, with confidence intervals, and false-positive rates per study with exact intervals. (D) Stage 1 test-set metrics for the full test set, the within-study subset, and the panels plus mutant-only studies: the all-mutant baseline, balanced accuracy, specificity with Clopper--Pearson CI, and AUC. Group values \texttt{panel-control} and \texttt{within-study-control} denote strain-panel and within-study controls, and ``panels'' denotes the strain panels. (E) Stage 1 threshold sweep by provenance: the four provenance groups per partition, specimens called mutant at each threshold with exact intervals, validation-chosen operating points, and the step-function curves behind \cref{fig:MS-SI-FIG-THRESHOLD-SWEEP}.

\item\phantomsection\label{tab:MS-SI-WB-BASELINES}\textbf{\siname{MS-SI-WB-BASELINES}. Baseline classifiers on the existing splits and the matched cohort. Related to \cref{fig:MS-FIG-POOLED-DESIGN}.} Sheets: (A) Baseline classifiers on the existing splits (per subset and stratum, with intervals) and the matched-cohort per-fold, per-stratum, and leave-one-study-out results. (B) Landmark availability per structure and study. Mandible-only, endocast-only, cranium-and-mandible, and all-three-structure landmark classifiers on the existing splits (within stratum and pooled, with the pooled Stage 2 error decomposition) and on the matched cohort (per stratum, aggregate, and leave-one-study-out), with the CNN and majority-class baselines re-scored on the same specimens. Per-structure component counts and whether the component cap was reached for every fit. (C) Non-CNN baselines on the existing splits: metadata-only, centroid-size, and landmark classifiers beside the CNN, Stage 1 by provenance subset and Stage 2 within each stratum, with the majority-class baselines. In sheets A and B, \texttt{lda} (also \texttt{gm-lda}, \texttt{landmark\_lda}) denotes the shrinkage (Ledoit--Wolf) linear discriminant, the primary landmark classifier, and \texttt{logistic} (also \texttt{gm-logistic}, \texttt{landmark\_logistic}) the L2-regularized logistic regression, the secondary classifier (\methodsref{sec:baseline-classifiers}). Stage 2 strata are abbreviated as Spry (all strains), Enhancer-all (all strains), Enhancer-B6 (C57BL/6J), Enhancer-FVB (FVB), and Enhancer-B6;129 (C57BL/6J;129).

\item\phantomsection\label{tab:MS-SI-WB-MATCHED-COHORT}\textbf{\siname{MS-SI-WB-MATCHED-COHORT}. Matched cohort. Related to \cref{fig:MS-FIG-MATCHED-CNN}.} Strata are study $\times$ strain $\times$ life phase (16 in the cohort), and the task within each is control-versus-mutant classification. Study names (for example Bmp) are MusMorph \texttt{Dataset} values, not gene labels. Sheets: (A) Matched cohort: exclusion steps, dropped strata with reasons, per-stratum covariate composition before and after balancing, membership in the existing splits, split checks, and the metadata-only balance check for the cohort and for the variant stratified on study and strain only. (B) Matched-cohort CNN: run log, train-vs-test balanced accuracy per run, per-repeat temperature and calibration, per-stratum and per-study AUC at each seed with seed stability, and the size-direction table (standardized difference in mean centroid size between mutants and controls in each stratum). (C) Classifiers on the matched cohort: within-stratum and cross-study aggregates. (D) Paired CNN$-$baseline differences in AUC for each seed and baseline, and in balanced accuracy for the cranium-and-mandible classifier: mean over the 16 strata with a bootstrap over strata (two forms) and a repeat-level $t$-interval on four degrees of freedom; counts of positive strata.

\end{itemize}